\documentclass[twocolumn,showpacs,preprintnumbers,amsmath,amssymb,prl]{revtex4}
\usepackage{graphicx}
\usepackage{graphicx}
\usepackage{epstopdf}

\begin{document}
\title{Critical phenomena  in one dimension from a  Bethe ansatz perspective}

\author{ Xiwen Guan}
\email{xiwen.guan@anu.edu.au}
\affiliation{State Key Laboratory of Magnetic Resonance and Atomic and Molecular Physics, Wuhan Institute of Physics and Mathematics, Chinese Academy of Sciences, Wuhan 430071, China}
\affiliation{Department of
Theoretical Physics, Research School of Physics and Engineering,
Australian National University, Canberra ACT 0200, Australia}

\date{\today}

\begin{abstract}
This article briefly reviews recent theoretical developments in quantum critical phenomena in one-dimensional (1D) integrable quantum gases of cold atoms. 
We present a discussion on quantum phase transitions, universal thermodynamics, scaling functions and correlations for a few prototypical exactly solved models, such as the Lieb-Liniger Bose gas, the spin-1 Bose gas with antiferromagnetic spin-spin interaction, the two-component interacting Fermi gas as well as spin-3/2 Fermi gases.
We demonstrate that their corresponding Bethe ansatz solutions provide a precise way to understand quantum many-body physics, such as quantum criticality, Luttinger liquids, the Wilson ratio, Tan's Contact, etc. 
These theoretical developments give rise to a physical perspective using integrability for uncovering experimentally testable phenomena in systems of interacting bosonic and fermonic ultracold atoms confined to 1D.

 \end{abstract}

\pacs{03.75.Ss, 03.75.Hh, 02.30.Ik, 05.30.Rt}
\maketitle

\section{I. Introduction}

\subsection{Critical phenomena}

Critical phenomena  are found everywhere  in nature, ranging from classical  phase transitions driven by thermal fluctuations to  quantum phase transitions driven by quantum fluctuations. 
Quantum critical phenomena  describe universal  scaling laws  of thermodynamic  properties for quantum many-body systems near a  phase transition at low temperatures. 
In general, quantum fluctuations couple strongly with thermal fluctuations near a quantum critical point.
When the thermal energy $k_BT$  is  less  than the energy gap $\Delta $,  physical  quantities  are dominated by quantum fluctuations.  Here $k_B$ is the Boltzmann constant.
On the other hand,   when $k_BT$ is larger than the energy gap, thermal fluctuations  dominate the order parameter fluctuations and control the critical behaviour. 
The major focus of quantum criticality  are the critical exponents and  the universal scaling functions  which control the thermodynamics of the critical matter between two stable phases near a critical point. 
Understanding criticality  is  still among the most challenging problems in condensed matter physics  \cite{QC-Book,Coleman}.

For a second order  quantum phase transition, the critical behaviour near the critical point is characterized by  a divergent correlation length $\xi \sim |g-g_c|^{-\nu}$ and the energy gap $\Delta$, which vanishes inversely proportional to the correlation length as $\Delta \sim \xi^{-z}  \sim |g-g_c|^{z \nu}$.  
Here the dynamical critical exponent $z$ and correlation length exponent $\nu$ are universal
and  $g$ is a  driving parameter, such as the chemical potential, magnetic field, interaction strength, etc.  
The dynamic critical exponent $z$ characterizes  the typical time scale of how the energy gap approaches  zero, i.e. $\tau _c \sim \xi^z$. 
This leads to  a phase coherence in the quantum critical regime. 
In this critical region,  a universal scale-invariant description of the system is expected through the  thermodynamic  properties.  
These divergences present novel critical phenomena \cite{QC-Book,Fisher,Vojta}. In  this regard, the critical temperature of the  quantum phase transition is $T_c=0$. 

 \begin{figure}[t]
  \includegraphics[width=0.790\linewidth,angle=90]{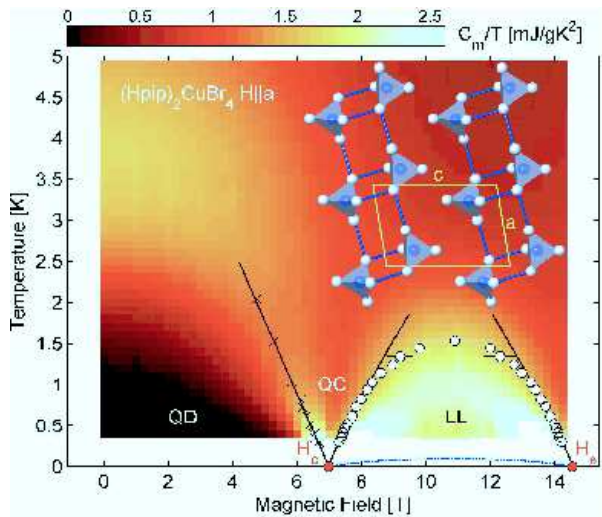}
\caption{The temperature scaled specific heat is plotted as a function of magnetic field and temperature for the quasi-1D strong coupling spin-ladder compound (Hpip)${}_2$CuBr${}_4$ \cite{ladder3}. The quantum phase transition is driven by the  external magnetic field.   The contour plot  indicates different phases   of the compound at low temperatures, i.e.   the quantum disorder phase (QD), quantum critical (QC) regime and spin Luttinger liquid (LL) phase. Circles denotes the LL crossover temperature which separates  the LL phase from the QC phase \cite{ladder3}.  The single  LL phase   describes the low energy physics of the  gapless regime.  The crossover temperature is proportional to $|H-H_c|$. This   marks a universality class with dynamic exponent  $z=2$ and correlation length exponent $\nu=1/2$ \cite{Masaki}.  Figure extracted from  \cite{ladder3}. }
\label{fig:ladder}
\end{figure}

One can  understand  general features of quantum criticality from the relation between quantum  and classical phases transitions. 
 For   a continuous  quantum phase transition, it is usual  to an  introduce imaginary time $\tau =1/k_BT$ to act as an additional space dimension so that as $T\to 0$, $\tau\to \infty$, i.e. 
  the extension of the system in this direction is infinite as the temperature tends to zero. 
 The time scale relates  the $z$th order of correlation length via $\tau_c \sim \xi^{-z}  \sim  |t|^{z \nu}$ with $t=|T-T_c|/T_c$. 
 Thus the quantum phase transition in $d$ space dimensions is related to   a  classical phase transition in $d+z$ space dimensions, see reviews \cite{QC-Book,Vojta}. 
 Both classical and quantum critical behaviour  are governed by divergent correlation lengths.
 Phase transition driven by quantum fluctuation, such as the $\lambda$-transition occur at finite temperatures  $T_c$, however, their   critical behaviour  is treated as  classical.  
In such  case, the finite temperature $T_c$ phase  boundaries  separate thermally disordered  and  ordered phases from the quantum critical region.


One  remarkable feature of  criticality is  universality. 
The  universality class of quantum phase transitions is classified by critical exponents  that  solely depend on symmetry of the excitation spectrum and dimensionality.  
The critical exponents are the same for systems in the same universality class. 
There are various methods  used to study critical behaviour of quantum many-body systems,  
such as the renormalization group approach \cite{Wison,Senthil}, continuum field theory and  the quantum-classical  mapping method \cite{QC-Book}.  Understanding quantum criticality and scalings  in quantum many-body systems still imposes formidable theoretical difficulties. 
It is therefore   highly desirable to present an exact  treatment  of  quantum phase transitions. 
To this end, exactly solvable models of cold atoms, of the strongly correlated electrons and spins that  exhibit  quantum phase transitions  provide  a rigorous way to  treat  quantum  phase transitions \cite{Masaki,Guan-Ho,GB}. 

\begin{figure}[t]
{{\includegraphics [width=1.00\linewidth,angle=-0]{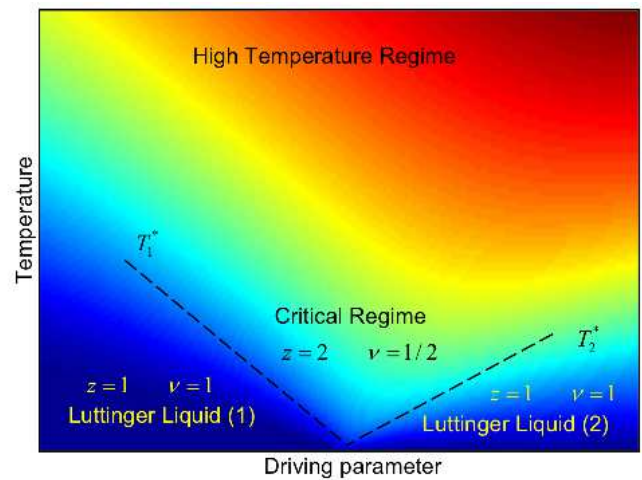}}}
\caption{Universal quantum criticality of a 1D quantum gas for the quantum phase transition from one ground state  to another.  The universal scalings of thermodynamics  with a finite temperature asymptotic  correlation length $\xi \sim 1/\sqrt{T}$ prevails in the critical regime, i.e.   $T> T^{*}_1$ and $T>T_2^{*}$. Here $T^*$ is proportional to the gap $|\mu-\mu_c|^{\nu z}$.
For the temperatures below crossover temperatures, i.e. $T< T^{*}_1,\,T^{*}_2$, two  different relativistic LLs with  exponent $z=1$ and correlation exponent $\nu=1$ display a  correlation length  $\xi \sim v/T $ in these LL regimes.  The high-temperature regime does not exhibit universal critical behaviour. }
  \label{fig:LL}
\end{figure}


\subsection{Quantum criticality in one dimension}

It is well known that 1D quantum systems exhibiting diverse  phase transitions are a rich resource to study quantum criticality.
In particular, the  quasi-1D spin ladder models  present rich  quantum phase transitions induced by external magnetic fields  \cite{Dagotto}. 
Quantum  criticality of 1D spin ladders is intimately related to
the  the SU(N) spin models with either integer  or half odd integer spins.  
For even-leg spin ladders, the phase transition between 
the gapped and gapless phases is  driven by the  external magnetic field. 
The nature of the gapless phase  illustrates 
the microscopic origin of criticality, see Fig.~\ref{fig:ladder}.
When the spin gap  is closed  by an external field,  the massive magnons of spin-$0,\pm 1$ states are restrained onto spin-$1$ states with a  non-relativistic dispersion relation \cite{Masaki}, $E(k)\approx \Delta +k^2/2m^* -HS^z$.  Here $m^*$ is the effective mass of magnons,  $\Delta$ is  the energy gap and $H$ is the external magnetic field. 
Near the critical point,  the $S^z=0,-1$  magnons  are not populated  due to the strong magnetic field. 
Thus the $S^z=1$  magnons can be mapped onto the free-fermion  universality class with  dynamics exponent $z=2$ and and correlation length exponent $\nu=1/2$.  
For the temperature below a crossover temperature $T^{*}\sim  |H-H_c|$,  a single-component Luttinger liquid (LL) with linear dispersion is presented in  the  whole gapless regime, i.e.  $H_c<H<H_s$.  
Here $H_c$ and $H_s$  are the lower and upper  critical fields, respectively. 
The physics of the gapless phase 
in one dimension is universally described by the Luttinger liquid with algebraic decay of the spin correlations at zero temperature, 
see  experiments \cite{ladder1,ladder2,ladder3,ladder4,ladder5,Watson}.  
For temperature $T>0$, the correlation functions in LL phase decay exponentially. 
The LL  parameters can be controlled 
 directly by the external magnetic field. 
In a recent insightful  experiment  \cite{Ninios2012},  this LL is shown to   display a  Fermi liquid nature at a renomalization fixed point.  
 This marks an intrinsic connection between the  two low energy theories: LL  physics in 1D and Fermi liquid  in higher dimensions. 
In contrast, spin triplet excitations in spin ladders in higher dimensions
can be described as Bose-Einstein condensed (BEC) magnons \cite{Giamarchi2}. 


In most of the integrable models, the eigenvalues can be obtained by means of the  Bethe ansatz wave function. 
The Bethe ansatz wave function  actually converts  the  Schr\"{o}dinger equation of a many-body problem   into a set of algebraic equations which are called the Bethe ansatz equations. 
These algebraic equations describe   the  roles of individual particles  in the presence of many others.
 The summation of such complex  individual roles often    leads to a global coherent state, i.e.   excitations  can form a collective motion of spin and   charge density waves  with different velocities (behave like bosons)  which are called the LL in spin and charge degrees of freedom.  

 These  studies  show that  low temperature thermodynamics of the  LL in the quantum disordered phase  and 
collective thermal  excitations  in quantum  critical regime have  significantly different critical behaviour, see Fig.~\ref{fig:LL}.
The collective  behaviour  in the quantum critical regime is determined by the thermal excitations of the  ground state.
Modifying  the  Yang-Yang grand canonical thermodynamic  approach \cite{Yang-Yang}, the  equation of states of 1D many-body systems can be obtained in entire  physical regimes.  
However, the Yang-Yang grand canonical description of the thermodynamics  of  integrable models is  always much involved  due to  the complexity of microscopic roles.
A new approach to treat thermodynamics of integrable systems  \cite{Erhai,Guan-Ho,GB} allows one to capture essential many-body physics in a rigorous way, including 
quantum criticality and quantum correlations.  

The 1D spinless Lieb-Liniger gas \cite{Lieb-Liniger,Cazalilla} and  the spin-1/2 Fermi gas \cite{Yang,Gaudin,Guan-RMP} are  among the most extensively studied integrable  many-body systems in quantum statistical mechanics. 
They exhibit novel quantum critical phenomena, and  have  had tremendous impact in quantum statistical mechanics. 
 The  quantum criticality of these models  involves a universal crossover from the relativistic LLs with  dynamic  exponent $z=1$ and correlation exponent $\nu=1$ to  free fermion quantum criticality with the dynamic exponent $z=2$ and correlation exponent $\nu=1/2$, see Fig.\ref{fig:LL}. 
  In this figure, the dashed lines indicate  crossover temperatures which separate the  quantum critical regime with a nonrelativistic dispersion from a relativistic LL phases at low temperatures. 
 In the critical region,  both correlation length and thermal wave lengths are proportional to $1/\sqrt{T}$.  Therefore, thermal and quantum fluctuations couple strongly at quantum criticality. 
 %
 %
In this scenario, recent breakthroughs in the experiments on trapped ultracold  bosonic and fermionic atoms confined 
to one dimension  has  provided  a better understanding of significant quantum statistical effects and quantum  correlations  in  many-body  systems, see  review \cite{Cazalilla,Guan-RMP}. 
In the present  review, we  briefly discuss    quantum criticality and the universal nature of quantum liquids for these archetypical  solvable models, i.e. Lieb-Liniger Bose gas, spin-1 Bose gas, the Gaudin-Yang model and large spin Fermi gases. 

The paper is organised as follows.   In Sec.~II, we discuss  quantum criticality and universal thermodynamics of the Lieb-Linger model. 
In Sec.~III, quantum critical behaviour of  the 1D spin-1 Bose gas with antiferromagnetic spin-spin interaction is discussed in terms of the grand canonical ensemble.  
In Sec.~IV we  discuss many-body critical phenomena and the  Fermi liquid nature  in the Yang-Gaudin  model. 
%
%
Section V presents an outlook on quantum criticality of large spin Fermi gases  and discusses new trends in   exactly solvable systems.


\section{II. Quantum criticality of Bose gases}

\subsection{Lieb-Liniger model}

The 1D Lieb-Liniger model \cite{Lieb-Liniger} of interacting bosons is a prototypical   Bethe ansatz  integrable model, see the review  \cite{Cazalilla,Korepin,Takahashi-b}. 
The   model  is described by the Hamiltonian
\begin{equation}
{\cal H}=-\frac{\hbar ^2}{2m}\sum_{i = 1}^{N}\frac{\partial
^2}{\partial x_i^2}+\,g_{\rm 1D} \sum_{1\leq i<j\leq N} \delta (x_i-x_j)
\label{Ham-1}
\end{equation}
in which $N$ spinless bosons, each of mass $m$, are constrained by periodic boundary conditions
on a line of length $L$ and $g_{\rm 1D} ={\hbar ^2 c}/{m}$ is
an effective one-dimensional coupling constant with scattering strength $c=-2/a_{1D}$ in a quasi-1D confinement. 
Here  $a_{1D}=\left( -a_{\perp }^{2}/2a_{s}\right) \left[ 1-C\left(
a_{s}/a_{\perp }\right) \right] $ is the 1D  scattering length with $a_{\perp }=\sqrt{2\hbar /m\omega _{\perp }}$ and the numerical constant $C\approx 1.4603$ \cite{Olshanii}. 
The dimensionless interaction strength is defined by $\gamma =c/n$,  where $n=N/L$ is the linear density.

Experimental studies of this model by using  cold bosonic atoms over a wide range of tunable interaction strength between atoms have  demonstrated the unique beauty of  the Bethe ansatz integrability, see a feature review article \cite{Batchelor-exp}.
These include the ferminization of the  Tonks-Girardeau gas \cite{Paredes,Kinoshita2004,Kitagawa}, quantum correlations \cite{Kinoshita2005,Guarrera}, thermolization \cite{Kinoshita2006},  Yang-Yang thermodynamics \cite{Exp4,Vogler},  the super Tonks-Girardeau gas \cite{Exp7},  quantum phonon  fluctuations \cite{Armijo,Exp-12},  elementary excitations \cite{Fabbri},  etc. There are more   experimental developments with the Lieb-Liniger model, see review \cite{Cazalilla,Guan-RMP}. 

Yang and Yang \cite{Yang-Yang} introduced the  particle-hole grand ensemble
to describe the thermodynamics of the model in equilibrium, which is later called  the thermodynamics Bethe ansatz (TBA) method by Takahashi  \cite{Takahashi71-74}. 
At finite temperatures the equilibrium states become degenerate. Yang and Yang showed that  true physical states can be determined by  the minimisation conditions of Gibbs free energy subject to constraints on the roots of the Bethe ansatz equations, see a commentary in \cite{Yang-a}. 
In terms of the dressed energy $\epsilon(k) = T\ln(\rho^h(k)/\rho(k))$ determined by the  particle density $\rho(k)$ and hole density $\rho^h(k)$  with  respect to 
the quasimomentum $k$ at finite temperature $T$, the Yang-Yang  equation is given by 
\begin{equation}
\epsilon (k)=
\epsilon^0(k) -\mu-
\frac{T}{2\pi}\int_{-\infty}^{\infty} dq\, a_2(k-q) \ln(1+{\mathrm e}^{-\frac{\epsilon(q)}{T}}) 
 \label{TBA-LL}
\end{equation}
where $\mu$ is the chemical potential,  $\epsilon^0(k)=\hbar^2k^2/(2m)$ is the bare dispersion and kernel function $a_n(k)=\left(nc/(2\pi)\right)/\left((nc/2)^2+k^2\right)$.
The dressed energy $\epsilon(k)$ plays the role of excitation energy
measured from the energy level $\epsilon(k_{\rm F})=0$, where $k_{\rm
F}$ is the Fermi-like momentum. 
The pressure $p(T)$   are given in terms of the dressed energy by
\begin{eqnarray}
p(T)&=&\frac{T}{2\pi}\int_{-\infty}^\infty dk \, \ln(1+\mathrm{e}^{-{\epsilon(k)}/{T}}).\label{Pressure-LL}
\label{pressure}
\end{eqnarray}
Eq.~(\ref{TBA-LL}) and (\ref{Pressure-LL}) present  a grand  canonical description of Bethe ansatz equations for the integrable Lieb-Liniger Bose gas.  
 It turns out that  the Yang-Yang grand canonical description is an elegant way to analytically access thermodynamics, LL physics and quantum criticality  \cite{GB,Meishan}.   
 From the TBA equation (\ref{TBA-LL}), it is obvious that the pressure (\ref{Pressure-LL})  reduces to several limiting cases, i.e. 
 the classical Boltzmann gas with  the de Broglie wavelength $\lambda_T^{-1}= \sqrt{m k_BT/2\pi\hbar^2}$, free Fermi gas and free Bose gas
\begin{equation}
p= \left\{ \begin{array}{ll}
\lambda_T\,k_BT^{-2} \, \mathrm{e}^{{\mu}/{T}}& T\to \infty\\
-\lambda_T \, k_BT \, \mathrm{Li}_{\frac{3}{2}} (-\mathrm{e}^{{\mu}/{T}}) & c \to \infty\\
\lambda_T \, k_BT \, \mathrm{Li}_{\frac{3}{2}} (\mathrm{e}^{{\mu}/{T}})& c\to 0
\end{array} \right. .
\end{equation}
Where $\mathrm{Li}_n(x) = \sum_{k=1}^\infty x^k/k^n$ is  the standard polylogarithm function.
This result suggests that the TBA equation (\ref{TBA-LL}) encodes three  statistics, Boltzmann statistics, Fermi-Dirac statistics and Bose-Einstein statistics in different limits.

\begin{figure}
  \centering
  \scalebox{0.400}{\includegraphics{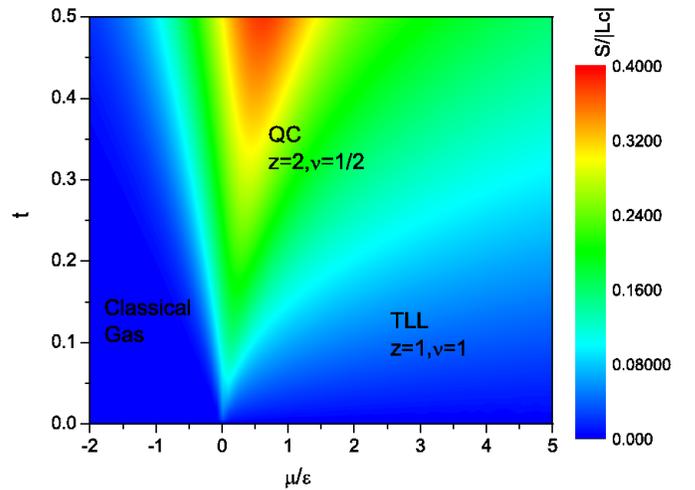}}
  \caption{Quantum phase diagram of Lieb-Liniger bosons at finite temperature: the contour plot of the entropy in
  the dimensionless  $t-\mu$ plane \cite{note1}. Here the temperature is rescaled $t=k_BT/\epsilon$. The critical point $\mu_c=0$.   At low temperature $T\ll |\mu-\mu_c|$ and $\mu<0$, it is a dilute classical gas. For temperature $T\ll \mu $ and $\mu>0$, the LL gives the low energy physics of the the system with  the dynamic exponent $z=1$ and correlation length exponent $\nu=1$. Near the critical point $\mu=0$ and the temperature $T\gg |\mu|$, it presents  a universality class of critical phase with $z=2$ and correlation length exponent $\nu=1/2$, also see \cite{GB}.  Figure extracted from \cite{Meishan}.  }
\label{fig:QC-LL}
\end{figure}

Although it  is generally believed that  there does not exist  a quantum phase transition at finite temperatures in this model,  for  grant canonical ensemble,  there  exists one critical point, i.e. chemical potential $\mu=0$, which separates the vacuum from the  filled ``Fermi sea" of bosons at zero temperature. 
It is shown that the TBA equation (\ref{TBA-LL}) becomes dimensionless in term of a rescaled temperature $t=k_BT/\epsilon$, here we defined the interaction energy  $\epsilon=\hbar^2c^2/(2m)$.  
At low temperatures, i.e. $k_BT \ll \epsilon$, the TBA equation (\ref{TBA-LL}) captures universal critical behaviour near $\mu_c=0$. 
This can be clearly seen from the finite temperature phase digram of this model which  was presented in see Fig.\ref{fig:QC-LL}.  Here entropy in $T-\mu$ plane presents a universality class of quantum criticality  associated quantum phase transition from vacuum into the LL phase. 
The right crossover temperature $T^{*} \sim |\mu-\mu_c|$ separates quantum critical regime with a nonrelativistic dispersion from the relativistic LL  with  exponent $z=1$ and correlation exponent $\nu=1$.  
The regime   $T\ll |\mu-\mu_c|$ and $\mu <0$ can be taken as a semiclassical gas, where  the de Broglie wavelength  $\lambda_T$ is much smaller than  the inter-particle mean spacing $1/n$ with   the density  $n \sim (1/\lambda )e^{-|\mu|/T}$.  In this ideal gas phase, there is no "Fermi surface".  The correlation behaves  different from the algebraic behaviour in the LL phase \cite{Gohmann:1999}.

Despite the equation (\ref{TBA-LL}) can be numerically solved for arbitrary strong interaction strength at all temperatures,  understanding universal nature of such archetypical many-body model acquires additional analytical  finite temperature thermodynamics. 
In \cite{GB}, Guan and Batchelor calculated the thermodynamics of the model  in analytic fashion using the polylog function.  The pressure can be cast into the form 
\begin{eqnarray}
p \approx -\sqrt{\frac{m}{2\pi\hbar^2}} T^{\frac{3}{2}} \, \mathrm{Li}_{\frac{3}{2}} (-\mathrm{e}^{A/T})
 \left[1-\frac{p}{\hbar^2c^3/(2m)}
 \right]
  \label{pressure-polylog}
\end{eqnarray}
where 
\begin{eqnarray}
A&=&\mu+\frac{2\,p(T)}{c}+\frac{1}{2\sqrt{\pi}c^3}\frac{T^{\frac{5}{2}}}{\left(\frac{\hbar^2}{2m}\right)^{\frac{3}{2}}} \mathrm{Li}_{\frac{5}{2}}
(-\mathrm{e}^{A_0/{T}}).\label{EoS-A}
\end{eqnarray}
Here $A_0\approx \mu+\frac{2\,p(T)}{c}$. This makes no  need to numerically solve the  integral equation (\ref{TBA-LL}) for critical phenomena at the temperatures below the degenerate temperature $T_F=\frac{\hbar^2}{2m}n^2$. 

\begin{figure*}[tbp]
\includegraphics[width=0.670\linewidth]{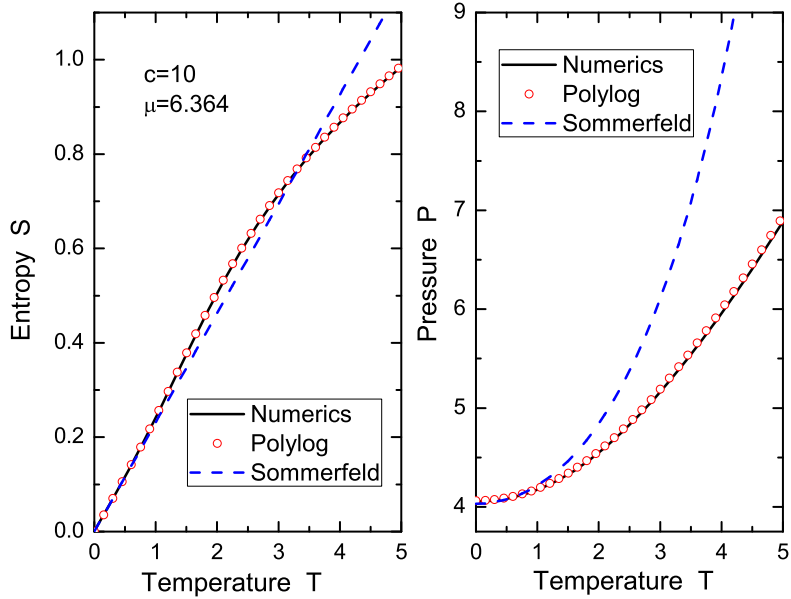}
  \caption{ Left panel: Entropy vs temperature for the Lieb-Linger model with $c=10$ and $\mu=6.634$ in dimensionless units. The derivation from the linear-temperature dependent entropy marks the  breakdown of the LL at a crossover temperature. The black solid line shows the numerical result from the TBA equations (\ref{TBA-LL}).  The red dots denote the analytical result derived from the equation of state (\ref{pressure-polylog}). The blue dashed line is the Sommerfeld expansion result in  the LL phase \cite{GB}. Right panel:  pressure vs temperature:  comparison between analytical and numerical results of the pressure obtained from  numerics, the polylog
  function and Sommerfeld expansion. 
  Figure extracted from \cite{GB}.}
\label{fig:S-Lieb}
\end{figure*}

Further more, to the  leading order, the free energy follows as  the field theory form 
\begin{equation}
F(T) \approx E_0-\frac{\pi C(k_BT)^2}{6\hbar v_c} \label{F-T}
\end{equation}
where the central charge $C=1$. For strong coupling regime,  $E_0$ is the ground state energy per length and the sound velocity $v_c$ are given by 
\begin{eqnarray}
E_0&\approx &\frac{1}{3}n^3\pi^2\left(1-\frac{4}{\gamma}+\frac{12}{\gamma^2}+\frac{\left(\frac{32}{15}\pi^2-32\right)}{\gamma^3}\right), \label{E}\\
 v_s&\approx & \frac{\hbar\pi n}{m} \left(1-\frac{4}{\gamma}+\frac{12 }{\gamma^2}+\frac{16}{\gamma^3}\left(\frac{\pi^2}{3}-2 \right)\right).
\end{eqnarray}
This implies that for temperatures below a crossover value $T^{*}$, the low-lying excitations have a linear relativistic dispersion relation,
i.e. of the form $\omega(k)=v_c(k-k_F)$. 
If the temperature exceeds this crossover value, the excitations involve free quasiparticles with
nonrelativistic dispersion. 
This crossover temperature can be
identified from the breakdown of linear temperature-dependent specific heat, i.e.  
\begin{equation}
S= \frac{\pi C(k_B)^2T}{3\hbar v_c},
\end{equation} 
see the Fi.~\ref{fig:S-Lieb}. 

In the quantum critical regime,  the density obeys the universal scaling form. 
It was proved \cite{GB} that the thermodynamic functions of this model such as the density and compressibility  can be cast into the 
universal scaling forms \cite{QC-Book,Fisher}
\begin{eqnarray}
n(\mu,T)&=&n_0+T^{\frac{d}{z}+1-\frac{1}{\nu z}}{\cal G}\left(\frac{\mu-\mu_c}{T^{\frac{1}{\nu z}}}\right),
\label{uni_scal_dens}\\
\kappa(\mu,T)&=&\kappa_0+T^{\frac{d}{z}+1-\frac{2}{\nu z}}{\cal F}\left(\frac{\mu-\mu_c}{T^{\frac{1}{\nu z}}}\right).
\label{uni_scal_kappa}
\end{eqnarray}
with dimensionality $d=1$. From the equation of state (\ref{pressure-polylog}),  the scaling function is 
\begin{equation} 
{\cal{F}}(x)=-\frac{c}{2\sqrt{\pi}}\mathrm{Li}_{\frac{1}{2}} (-\mathrm{e}^x)
\end{equation} 
 for $T> |\mu-\mu_c|$ which reads off  the background density   $n_0(T,\mu)=0$ and the critical exponent $z=2$ with  the correlation
length exponent $\nu=1/2$.
The free fermion criticality revealed from (\ref{uni_scal_kappa}) naturally comes from  the fact that near the critical point $\mu_c=0$ the system has low density and strong interaction, i.e. Tonks-Girardeau regime. 
Meanwhile,  the compressibility satisfies the scaling form (\ref{uni_scal_kappa})
with ${\cal{F}}(x)=-\frac{c}{2\varepsilon \sqrt{\pi}}\mathrm{Li}_{-\frac{1}{2}}(-\mathrm{e}^x)$ and the background density 
$\kappa_0 (T,\mu)=0$.  

 In the realistic  experiment, the quantum criticality (\ref{uni_scal_kappa}) can be mapped out from the density profile of the trapped gas at finite temperature. 
 Within the local density approximation,
the chemical potentials in the equation of state (\ref{pressure-polylog}) as well as in the TBA equation (\ref{Pressure-LL})
 are replaced by the local chemical potentials given by
\begin{eqnarray}
\mu\left( x\right) &=&\mu\left( 0\right) -V\left( x\right) ,
\label{LDA1} \label{LDA}
\end{eqnarray}%
Here the $\mu(0)$ is the trapping center chemical potential which can be fixed by the total number in the trap.  The external potential is defined as $V\left( x\right)
=m\omega ^{2}x^{2}/2$ with harmonic frequency $\omega $ and the
character length for the harmonic trap is $a=\sqrt{\hbar /m\omega
} $.
One can read off the  dynamic exponent $z=2$ and the correlation
length exponent $\nu=1/2$ from the  density curves with a proper temperature scaling  at  different temperatures \cite{Wang}.


Moreover,  the 1D integrable  critical   system, such as the LL,   gives  rise to the power law behaviour of  long distance or long time asymptotics of correlation functions.
Using the  conformal field theory (CFT), which preserves angles, enables one   to obtain asymptotic behaviour of correlation functions. 
 For   transformation $\omega=\omega(z)$ and $\bar{\omega}=\bar{\omega}(\bar{z})$, the primary fields transform as 
\begin{equation}
\phi(z,\bar{z})=\left(\frac{\partial \omega}{\partial z}\right)^{\Delta^+}\left(\frac{\partial \bar{\omega}}{\partial \bar{z} }\right)^{\Delta^-}\phi(\omega,\bar{\omega}).\label{CFT-t}
\end{equation}
Where $\Delta^{\pm}$ are two real numbers called the conformal weights of the field $\phi$. 
The correlation functions of these primary fields were given as \cite{Belavin:1984a,Cardy:1986}
\begin{equation}
\langle \phi (z_1,\bar{z_1})\phi(z_2,\bar{z}_2)\rangle =(z_1-z_2)^{-2\Delta^+}(\bar{z}_1-\bar{z}_2)^{-2\Delta^-}.
\end{equation}

If we set $z = v\tau  + \mathrm{i} y, \, \bar{z} = v\tau - \mathrm{i} y$  ($-\infty < \tau  <\infty,\,  -L \le  y \le 0$), where $\tau$  is the Euclidean time and $v$  is the velocity of light. The two-point correlation function  for primary fields with the conformal dimension $\Delta^{\pm}$  becomes 
\begin{equation}
\langle \phi (\tau,y)\phi(0,0)\rangle=\frac{\exp(2\mathrm{i}\Delta Dk_Fy)}{(v\tau +\mathrm{i} y)^{2\Delta^+}(v\tau -\mathrm{i} y)^{2\Delta^-}}. 
\end{equation} 
The conformal dimensions  $\Delta ^{\pm}$ can be read off from the finite-size corrections of low-lying excitations via Bethe ansatz solutions.
Moreover, by replacing $1/L$ with $T$ in the conformal map $z=\exp(2\pi\omega/L)$,  conformal invariance gives a universal form of correlation functions with exponential decay for a large distance asymptotic \cite{Cardy:1986,Affleck:1986}.
 The critical Hamiltonian can be approximately described  by the conformal Hamiltonian which is described by the generators  of underlying Virasoro algebra  and the central charge $C$. 
 %
 %
 For $C>1$, then the conformal dimensions characterizing the power law decay of the correlation functions at infinity depend continuously on the model parameters. 
 In particular, the LL   is specified by the central charge $C= 1$ Virasoro algebra, for examaple,   the Gaussian model 
\cite{Haldane:1981a,Haldane:1981b}.
Close to criticality,
the dispersion relations for 1D quantum systems are approximately
linear. 
Conformal invariance predicts that the energy per unit
length has a universal finite size scaling form $E=E_{0}+\Delta/L^2$
where $E_{0}$ is the ground state energy per unit length for the
infinite system and $\Delta$ is a universal term related the conformal weights.
Consequently, the calculation of the critical exponents of the critical models can reduce to calculation of low-lying excitation spectra. 
Thus the finite-size low-lying  corrections to the ground state energy essentially determine  the asymptotic behaviour of the correlation functions of the critical models, see \cite{Korepin,1D-Hubbard}.

\begin{figure}
  \centering
  \scalebox{0.8400}{\includegraphics{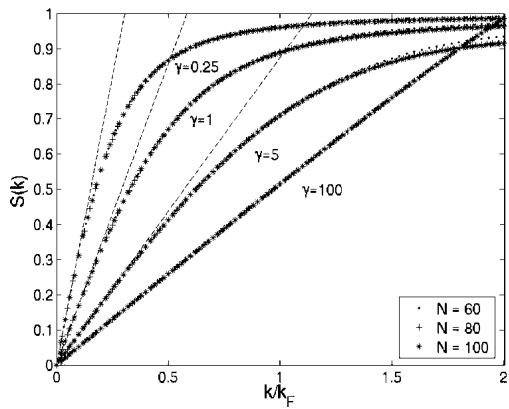}}
  \caption{Static ynamical structure factor for different interaction strength at temperature. The dashed lines show the small momentum asymptotic relation $S(0)=|k|/v_c$.    Figure extracted from \cite{Panfil:2014}.}
\label{fig:DSF}
\end{figure}

Although thermodynamics of the Lieb-Linger is accessible for  whole temperature regime, the dynamic density-density  correlation functions are difficult to calculate in analytical fashion \cite{CC}. Along  this line, Caux and his coworkers have devoted systematic study on zero and finite temperature dynamical correlations \cite{Caux:2006,Panfil:2014} based on the Bethe ansatz equations of the Lieb-Linger gas. The density-density correlation functions in Fourier space give  the dynamical structure factor 
\begin{equation}
S(k,\omega)=\int_0^Ldx\int dt e^{-\mathrm{i} kx+\mathrm\omega t}\langle \rho(x,t) \rho(0,0)\rangle
\end{equation}
with the density $\rho(x) =\sum_{j=1}^N\delta (x-x_j)$. At zero temperature, the dynamical structure factor can be represented as a sum of matrix elements of the density operator in the basis of Bethe eigenstate $\mid \alpha \rangle$, i.e.
\begin{equation}
S(k,\omega)=\frac{2\pi}{L}\sum_{\alpha}|\langle 0\mid \rho_k\mid \alpha\rangle  |^2\delta(\omega -E_\alpha +E_0)
\end{equation}
with $\rho_k=\sum_{j=1}^Ne^{-\mathrm{i} kx_j}$. The static structure factor is given by $S(k)=\int d\omega S(k,\omega)/2\pi$. At finite temperature, density-density function becomes
\begin{equation}
S(k,\omega)=\frac{2\pi}{L}\sum_{\lambda }|\langle \lambda_{\rho T}\mid \rho_k\mid \alpha\rangle  |^2\delta(\omega -E_\alpha +E_0),
\end{equation}
where $\mid \lambda_ {\rho T }\rangle$  is the thermal equilibrium state.  With the help of the  Bethe ansatz solution and f-sum rule, the static dynamical factor at zero was plotted in terms of the interaction strength, see Fig.~\ref{fig:DSF}.  The full dynamic correlations at finite temperatures have been studied recently in  \cite{Panfil:2014}.

On the other hand, due to linear dispersion, the correlation functions of the system can be calculated within Bosonization approach \cite{Giamarchi,Cazalilla,Cazalilla2}. 
The one-particle static  density matrix is given by 
\begin{eqnarray}
\langle \Psi^{\dagger}(x) \Psi(0) \rangle&=& \rho_0\left[\frac{1}{\rho_0d(x|L)} \right]^{\frac{1}{2K}}\left\{ b_0+\right. \nonumber \\
&& \left. \sum_{m=1}^{\infty} b_m \left[\frac{1}{\rho_0d(x|L) } \right]^{2m^2K}  \cos \left(2\pi m\rho_0x \right)\right\} \nonumber
\end{eqnarray}
Where the function $d(x|L)=L|\sin(\pi x /L)|/\pi$ and $b_m$ are dimensionless coefficients. $\rho_0$ is the linear density. 
 In the thermodynamics limit, $L\to \infty$, $d(x|L) \to |x|$. 
 $K$ is the Luttinger parameter which  determines the asymptotic behaviour of the correlation functions of the critical system. 
 In thermodynamic limit, the leading order of one-particle correlation $\langle \Psi^{\dagger}(x) \Psi(0) \rangle \sim  1/x^{1/2K}$. 
  One can observe that the oscillation terms become more important as the Luttinger parameter decreases. In strong coupling limit, the Luttinger parameter is given by \cite{Ristivojevic}
  \begin{eqnarray}
  K=1+\frac{4}{\gamma}+\frac{4}{\gamma^2}-\frac{16\pi^2}{3\gamma^3}+\frac{32\pi^2}{3\gamma^4}+O(\gamma^{-5}).
  \end{eqnarray}
  The  density-density  correlation function is given by 
\begin{eqnarray}
\langle \rho(x) \rho(0) \rangle&=& \rho_0^2\left\{ 1-\frac{K}{2\pi^2} \left[\frac{1}{\rho_0d(x|L)} \right]^{2} +\right. \nonumber\\
&& \left. \sum_{m=1}^{\infty} a_m \left[\frac{1}{\rho_0d(x|L) } \right]^{2m^2K}  \cos \left(2\pi m\rho_0x \right)\right\}. \nonumber
\end{eqnarray}
In the Tonks-Girardeau limit, the density correlation function reads
\begin{eqnarray}
\langle \rho(x) \rho(0) \rangle&=& \rho_0^2\left\{ 1-\left[ \frac{\sin(\pi\rho_0x)}{\pi\rho_0x}\right] \right\}.
\end{eqnarray}

 It is particularly interesting that the condensate fraction of the Lieb-Liniger model displays finite-size critical scaling behaviour \cite{J-Sato}.  
 At $T=0$, the Bose-Einstein condensation  for the 1D Lieb-Linger gas with repulsive short-range interactions  meets the Penrose and Onsager criterion by taking a particular large size limit.
 It is shown that if  the interaction strength $\gamma$ is given by a negative power of particle number $N$, i.e. $\gamma=A/N^{\eta} $, the condensate fraction $n_0:=N_0/N$ is nonzero and constant in various thermodynamic limits. 
 Here $A$ is a density-independent amplitude and $\eta$ is an exponent. 
  $N_0$ is the largest eigenvalue of the one-particle  reduced density matrix, which can be numerically evaluated by the Bethe ansatz equations \cite{J-Sato}.  Thus the Lieb-Liniger gas does show Boson-Einstein condensation in the sense of Penrose and Onsager criterion. 

Despite Lieb-Liniger is  the simplest integrable model, it has rich many-body phenomena and  gives deep insight into higher dimensional physics of many-body systems. 
The exact results for various observables of the Lieb-Liniger  model at T=0 and at finite temperature were obtained  by using field theory  methods, i.e. the  repulsive Lieb-Liniger Bose gas  can be obtained as the nonrelativistic limit of the sinh-Gordon model \cite{Kormos1,Kormos2}.
The local  two- and three-body correlations have been calculated from the powerful  field theory methods \cite{Kormos3,Kormos4}.
In particular, Haller et al. \cite{Exp7} ] made an experimental breakthrough by observing a metastable highly excited gas-like phase Ð called the super Tonks-Girardeau gas Ð in the strongly attractive regime of bosonic Cesium atoms. 
The super Tonks-Girardeau gas was predicted theoretically by Astrakharchik et al. \cite{Astrakharchik1,Astrakharchik2}  using Monte Carlo simulations and confirmed by Batchelor {\em et al.}  \cite{Batchelor-s}  using the integrable interacting Bose gas with attractive interaction. 
The study of transition dynamics from Tonks-Girardeau gas \cite{Girardeau} to attractive Tonks-Girardeau phase gives a further  understanding of such metastable states in many-body systems \cite{Chen:2010}. 
This gas-like excitation state can be viewed as more strongly correlated Tonk-Girardeau gas with metastable criticality \cite{Panfil:2013}. 
In that  paper, the authors proved that the  excited gas-like states  in the super Tonks-Girardeau gas are more favourable in certain strongly attractive interaction regime in comparison with the excitations of the cluster states of different sizes. 

\begin{figure}
  \centering
  {{\includegraphics [width=0.99\linewidth,angle=-0]{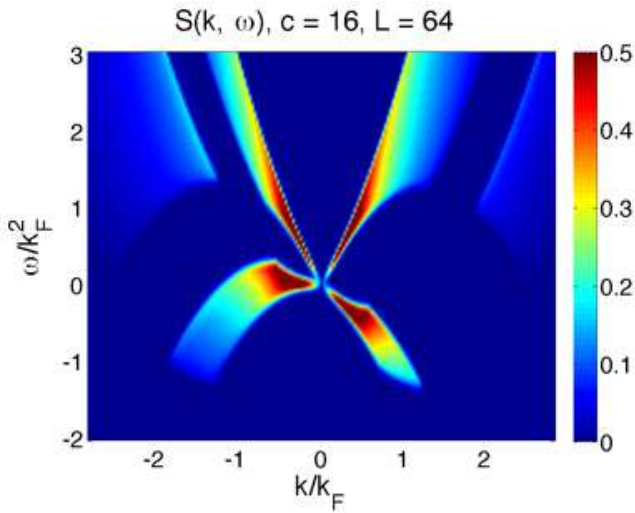}}}
  \caption{ Contour plot of dynamical structure factor for the Lieb-Liniger gas with two separated "Fermi seas" which was discussed in \cite{Fokkema}.
  The numerical calculation was set up with the interaction strength $c=16$ for total number $64$  bosons separated in two  asymmetric Fermi seas. 
  The effect of unbalanced Fermi seas is visualised.  It indicates the threshold singularity values are essentially momentum dependent regarding to the separating Fermi seas.  
  Figure extracted from  \cite{Fokkema}. }
\label{fig:Fermi-sea splitting}
\end{figure}

In regard to the highly excited state, the splitting up of the Fermi sea of the 1D Lieb-Lininger gas displays interesting properties associated with multiple Fermi momenta \cite{Fokkema}.
In this regard, the Fermi sea splitting leads to consequences on the spectrum of such highly excited states with multiple Fermi seas.
In particular, the dynamical structure factor displays threshold singularities at each edges of separated Fermi seas, see Fig.~\ref{fig:Fermi-sea splitting}.
The linearization of spectrum near each Fermi point also  gives rise to the effective LL description of the Fermi sea splitting phenomena. 

The Lieb-Liniger model provides a genuine setting to examine subtle many-body physics. 
The model presents the Tonks-Girardeau gas as the repulsion tends to infinity. 
In fact, the energy can be continuous in the limits $c\to \pm \infty$.  
 Such a novel connection at the $c\to \pm \infty$ results in a practicable quantum holonomy where the quantum states do not come back to the original ones after a cyclic changes of the interaction strength \cite{Yonezawa}.
 Beyond quantum holonomy, the complexification of the interaction $c$ and metastable quantum criticality of the supper Tonks-Girardeau gas \cite{Panfil:2013} will further stimulate study of quantum liquid phase in excitations and non-Hermitian systems where the linear dispersion spectrum can be robust.

\subsection{Spinor Bose gases}

Large spin Bose gases have a rich phase diagram  and exhibit    various phases of quantum liquids \cite{Ho} .
The physics of spin-1, spin-2 and  spin-3 bosons has been investigated in experiments with Na,  ${}^{87}$Rb and ${}^{52}$Cr cold atoms. 
Spinor ultracold gases with large spins  in one dimension  present novel ferromagnetism and various spin liquids  \cite{Cazalilla}.
The spinor Bose gases with spin-independent 
short range interaction have a ferromagnetic ground state, \cite{Eisenberg-Lieb,Yang-Li}. 
This nature was well  invested in the two-component  spinor Bose gas with spin-independent s-wave scattering  \cite{Sutherland,Li,GBT} and the integrable multi-component Bose gases \cite{Yang-You,Cao2}. 
At low energy level, the 1D two-component spinor Bose gas can be described by an effective Hamiltonian with spin-charge separation \cite{Matveev:2008,GBT}, i.e. $\mathcal{H} =\mathcal{H}_{\rm ph} +\mathcal{H}_{\sigma}$ with charge density  excitations and spin excitations
\begin{eqnarray}
\mathcal{H}_{\rm ph}  &=& \frac{\hbar u_\rho}{2\pi}\int\left( K(\partial_x \theta )^2+K^{-1} (\partial _x \phi )^2\right) dx \nonumber\\
\mathcal{H}_{\sigma} &=& -\sum _{\ell} J {\bf S}_\ell \cdot {\bf S}_{\ell+1}. 
\end{eqnarray} 
Where $u_\rho$ the sound velocity and $K$ is the Luttinger parameter   in charge sector.  
$\phi$ and $\theta$ are the bosonic fields of density and phase satisfying the communication relation $[\phi(x),\partial _y\theta (y)]=\mathrm{i} \pi \delta (x-y)$.  
The spin excitations can be described by the ferromagnetic Heisenberg chain with the effective coupling strength $J\approx 2P/c$ for strong coupling regime.
 Where  $P$ is the pressure of the gas. 
 In this strong coupling regime, i.e. $1\ll c/n \ll 1/k_BT$, 	the specific heat exponent for the spinor Bose	gas	behaves as $c_v \sim T^{1/2}$ which is different from the Lieb-Liniger gas for which $c_v \sim T$. 
 In long wave length limit, the spin wave excitations  over the ferromagnetic ground stat is given by $ \omega(p)=E-E_0=p^2/(2m^*)$ \cite{Fuchs,BBGO:2006}.
Here $m^{*}$ is the effective mass.  For strong coupling, it is given by 
\begin{equation}
\frac{m}{m^{*}}\approx \frac{1}{N}+\frac{2\pi^2 }{3\gamma }\left(1-\frac{2}{\gamma}\right).
\end{equation}
We see that  the effective mass takes the maximum value $m^{*} = Nm$ for $\gamma \to \infty$. 
This means  that by moving one boson with down spin, one has to move all the particles with up spins.

In contrast, the 1D  spin-1 bosons with  repulsive density-density and antiferromagnetic spin-exchange interactions  \cite{Cao,Shlyapnikov,Lee,Shlyapnikov2}  exhibit  either a spin-singlet paired ground state or a fully polarized ferromagnetic ground state.  
For the external field less than a lower critical field, the antiferromagnetic interaction leads to an effective attraction 
in the spin-singlet channel that gives rise to a quasi-condensate of singlet bosonic pairs. 
 In this phase, the 
low energy physics can be characterized  by a spin-charge separation theory of the $U(1)$ 
LL  describing the charge sector  and a $O(3)$ non-linear sigma model 
describing the spin sector \cite{Shlyapnikov}.
It was shown that this  model provides an  integrable regularization of the $O(3)$ nonlinear sigma model. 
On the other hand,  the solely ferromagnetic quasi-condensate of fully polarized  bosons  occurs  as  the  external field exceeds an upper critical field.
Exact Bethe ansatz solution of this  model   provides deep insights into understanding  competing ordering with  quantum criticality \cite{Kuhn,Wang-Zhang}.

The  Hamiltonian of the 1D spin-1 bosons with repulsive density-density and antiferromagnetic spin-exchange interactions  reads \cite{Cao}
\begin{eqnarray}
 {\cal H}=-\sum^N_{i=1}\frac{\partial^2}{\partial x^2_i}+\sum_{i<j}[c_0+c_2S_i\cdot S_j]\delta(x_i-x_j)+E_z,
\label{Ham-Bose-1}
\end{eqnarray}
that describes $N$ particles of mass $m$ confined in 1D to a length $L$ with 
$\delta$-interacting type density-density and antiferromagnetic spin-exchange interactions between two atoms.  
In the above equations,  $S_i$ is the spin-1 operator with $z$-component $(S^z=1, 0, -1)$. 
The interaction parameters  $c_0=(g_0+2g_2)/3$ and $c_2=(g_2-g_0)/3$ where the interaction strength is given by 
$g_S=4\pi\hbar^2a_S/m$. 
Here  $a_S$ represents the $s$-wave scattering length in the total spin $S=0,2$ channels. 
$E_z=-HS^z$ stands for the Zeeman energy, where $H$ is the external field and $S^z$ the total spin in the 
$z$-component. 
The model  (\ref{Ham-Bose-1}) has two conserved quantities, 
$S^z$ and the total particle number $N$. 
In this model, the number of particles in a particular spin state $(S^z= 1, 0, -1)$ is no longer conserved.

\begin{figure}[t]
{{\includegraphics [width=0.82\linewidth,angle=-90]{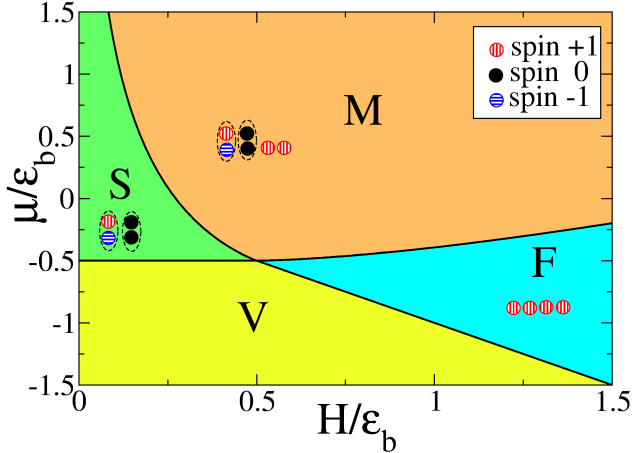}}}
\caption{(Color online) Phase diagram of the spin-1 Bose gas  in the $\mu$-$H$ plane \cite{Kuhn}.  There are three distinguishable phases:  the  spin-singlet phase $S$ of paired bosons,  
ferromagnetic phase $F$ of spin-aligned bosons, and  a mixed phase $M$ 
of  pairs  and unpaired bosons.  $V$ denotes   the vacuum. The solid lines show the  numerical result of the critical fields obtained from the TBA equations (\ref{TBA}) in $T=0$ limit. Figure extracted from \cite{Kuhn}.}
\label{fig:phase-spin-1} 
\end{figure}

The model with antiferromagnetic spin-exchange interaction for $c=c_0=c_2>0$  is exactly solvable  by the Bethe ansatz  \cite{Cao}. 
 In grand canonical ensemble, thermodynamics of the model  are determined by the the TBA equations (see \cite{Lee} for details).
In terms  of the dressed energies $\varepsilon_1(k)$, $\varepsilon_2(k)$ and $\phi_n(k)$ for unpaired states, paired states and spin strings, the TBA equations read
\begin{eqnarray}
\varepsilon_1(k) &=& k^2-\mu-H-Ta_4*\ln(1+e^{-\frac{\varepsilon_1(k)}{T}})\nonumber \\ 
&&+ T[a_1-a_5]*\ln(1+e^{-\frac{\varepsilon_2(k)}{T}}) \nonumber \\ 
&&- T\sum_{n=1}^{\infty}[a_{n-1}+a_{n+1}]*\ln(1+e^{-\frac{\phi_n(k)}{T} }), \nonumber \\ 
\varepsilon_2(k)&=& 2(k^2-c'^2-\mu)+T[a_1-a_5]*\ln(1+e^{-\frac{\varepsilon_1(k)}{T}})\nonumber\\
&&+ T[a_2-a_4-a_6]*\ln(1+e^{-\frac{\varepsilon_2(k)}{T} }), \nonumber\\ 
\phi_n(k)&=&n+T[a_{n-1}+a_{n+1}]*\ln(1+e^{-\frac{\varepsilon_1(k)}{T}}) \nonumber \\ 
&&+ T\sum_{n=1}^{\infty}T_{mn}*\ln(1+e^{-\frac{\phi_n(k)}{T}})
\label{TBA}
\end{eqnarray}
with $n=1,2,\ldots, \infty$. Here   the symbol $*$  denotes the convolution $(f*g(x))=\int_{-\infty}^{\infty}f(x-x')g(x')dx'$, the functions $a_n$ are  the same as the above and $T_{nm}$ was  given in \cite{Lee}.  

The phase diagram of the model (\ref{Ham-Bose-1}) presents a significant understanding of quantum phase transitions induced by any driving force, such as external fields, chemical potential, density and interaction strength. 
Exact solution of the model does provide precise determination of the phase diagram in an analytical way. 
The full phase diagram is presented  in $\mu-H$ plane, see Fig. \ref{fig:phase-spin-1}.
The model exhibits three quantum phases at zero temperature: spin-singlet paired bosons $S$, ferromagnetic spin-aligned bosons $F$, and a mixed phase of the pairs and unpaired bosons $M$. $V$ stands for the vacuum.
The spin-singlet paired  phase  involves two types of pairs:  pairs with different spin states $|F = 1,m_F=\pm 1 \rangle$  and   pairs of two $|F = 1, m_F=0\rangle $ bosons. 
The phase boundaries  in the $\mu-H$ plane are essential for determining the  quantum criticality of  the model. 

\begin{figure}[t]
{\includegraphics [width=1.0\linewidth]{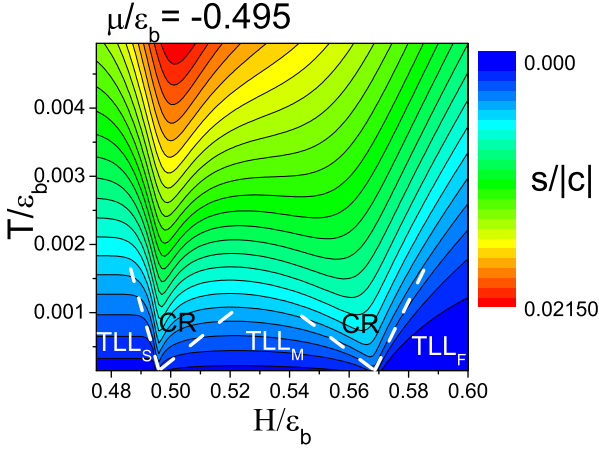}}
\caption{(Color online) Contour plot of the entropy $s$ vs the external field $H$ for  fixed  chemical potential $\tilde{\mu}=-0.495$  in the $T-H$ plane. 
The dashed lines indicate the crossover temperatures which are determined by the Luttinger liquid behaviour  of linear-temperature-dependent entropy (\ref{S-spin-1}).
These present   the universal crossover from the relativistic Luttinger liquid to the free fermion criticality with non-relativistic dispersion. Figure extracted from \cite{Kuhn}. }
\label{contourH} 
\end{figure}

The model exhibits the ground state of either spin-singlet pairs  or ferromagnetic fully-aligned bosons or co-existence of pairs and single bosons. %
The driving force like chemical potential and  external field can drive the system from one ground state to another so that the associated quantum phase transition leads to universal critical phenomena.  
In canonical ensemble, the system presents a gapped phase for the external field less than a lower critical field $h_{c1}$.  
For the external field excesses a upper critical field $h_{c2}$,  the fully- spin-aligned bosons form a ferromagnetic ground state. 
for an intermedium field, i.e. $h_{c1} \le h \le h_{c2}$, the ground state gives  the mixture  of the pairs and single bosons of state $|F = 1,m_F= 1 \rangle$.  
For strong coupling region the critical fields  read
\begin{eqnarray}
h_{c1}&=& -\tilde{\mu} + \frac{32\sqrt{2}}{15\pi}\left(\tilde{\mu}+\frac{1}{2}\right)^{\frac{3}{2}} - \frac{32}{45\pi^2}\left(\tilde{\mu}+\frac{1}{2}\right)^2,\nonumber  \\
h_{c2}&=& -\tilde{\mu} + \frac{1}{2}\left(\frac{15\pi}{4}\right)^{\frac{2}{3}}\left(\tilde{\mu}+\frac{1}{2}\right)^{\frac{2}{3}} - \frac{5}{8}\left(\tilde{\mu}+\frac{1}{2}\right).\nonumber
\end{eqnarray}
The quantum phase transitions driven by a magnetic field provide insight into large spin magnetism and universal criticality of the model.   
At low  temperatures, the 
three zero temperature  quantum phases, i.e.,  the phase of singlet pairs,  ferromagnetic phase of spin-aligned  atoms and the mixed phase 
of pairs and single atoms,  could form the relativistic  LL of bound pairs ($LL_S$), LL of  single atoms ($LL_F$) and a two-component LL  ($LL_M$) of 
paired and single atoms, respectively. 
This Luttinger liquid nature is evidenced from the universal form of the entropy
\begin{eqnarray}
s=\left\{  \begin{array}{ll}
\frac{\pi T}{3}\frac{1}{v_2},&{\rm for\, phase }\, S\\
\frac{\pi T}{3}(\frac{1}{v_1}+\frac{1}{v_2})&{\rm for\, phase }\, M\\
\frac{\pi T}{3}\frac{1}{v_1} &{\rm for\, phase }\, F
\end{array}
\right. \label{S-spin-1}
\end{eqnarray}
where the velocities can be calculated  by the TBA  equations (\ref{TBA}) for full interaction strength. 
For strong coupling limit, one can obtain explicit forms of the velocities
\begin{eqnarray}
v_1 &=& 2\pi n_1\left(1+\frac{2(32n_2-10n_1)}{5|c|}\right)\nonumber \\
v_2 &=& \pi n_2\left(1+\frac{2(48n_1+5n_2)}{15|c|}\right).\nonumber
\end{eqnarray}
Beyond  crossover temperatures $T^{*}\sim |h-h_c|$,  the low energy  LL physics  breaks down, i.e.  the dispersion of either bound pairs or unpaired single atoms becomes nonrelativistic.
These critical fields  and the crossover temperatures  can be worked out in a straightforward way. 
These are obviously evidenced  in the contour plot of  the entropy in the $T-H$ plane, see  Figure \ref{contourH}. 
The criticality  of the spin-1 Bose gas are described by the free fermion criticality of  two Tonks-Girardeau gases with masses $m$ and $2m$.

In the vicinity of the quantum critical points $h_{c1}$  and $h_{c2}$, the system exhibits two different quantum critical regimes. The scaling functions of density and compressibility  in the critical regime near the critical points $h_{c1}$ and $h_{c2}$ can be cast into the universal form (\ref{uni_scal_kappa}) with the driving parameter of $h$ instead the chemical potential $\mu$, see \cite{Kuhn}
Nevertheless, all thermodynamical properties like  magnetization and susceptibility associated with the phase transitions driven by the magnetic field were rescaled to the universality class of quantum criticality of free fermions.
It turns out that these universal  thermodynamical  properties can be used to map out the bulk phase diagram through the 1D  trapped gas at finite temperatures \cite{Kuhn}. 
By reformulating the equation of state  within the local 
density approximation (LDA)  with  a replacement $\mu(x)=\mu(0)-\frac{1}{2}m\omega_x^2x^2$ 
in which $x$ is the position and $\omega_x$ is the trapping  frequency, the density profile and thermodynamical properties can be obtained for the  trapped gas.
%

 

\section{IV. Probing Wilson ratio  and Contact  with criticality}

\subsection{Critical phenomena of the attractive Fermi gas}

The 1D $delta$-function interacting  spin-1/2 Fermi gas is a prototypical exactly solved model in literature. 
This model exhibits rich physics of interacting fermions in 1D: from few-body physics to many-body phenomena, including polarons, FFLO-like pairing, LL, quantum criticality, Fermi liquid signature, see a recent review \cite{Guan-RMP}. 
Here we briefly discuss how exact solutions provide insights into  the universal feature of Wilson ratio \cite{Wilson:1975}  and Tan's contact \cite{Tan}.

 In order to accommodate a general quantum liquid theory for the non-spin-charge separated mechanism, we briefly introduce a two-component attractive Fermi gas which was historically referred as Yang-Gaudin  model \cite{Yang:1967,Gaudin:1967}. 
 The quantum many-body  Hamiltonian  describes $N=N_{\uparrow}+N_{\downarrow}$ fermions of mass $m$ 
 with external magnetic field $H$ 
\begin{eqnarray}
\mathcal{H} &=&-\frac{\hbar ^{2}}{2m}\sum_{i=1}^{N}\frac{\partial ^{2}}{
\partial x_{i}^{2}}+g_{1D}\sum_{i=1}^{N_{\uparrow}}\sum_{j=1}^{N_{\downarrow }}\delta \left( x_{i}-x_{j}\right)+ E_z \quad
\label{Hamiltonian}
\end{eqnarray}
in which the  Zeeman
energy is given by $E_z= -\frac{1}{2}g \mu_BH\left( N_{\uparrow }-N_{\downarrow }\right)$. 
As being mentioned before, the effective 1D interaction strength  $g_{1D}=-2\hbar^2/(ma_{1D})$ can 
be tuned from the weakly interacting regime ($g_{1D}\rightarrow 0^{\pm}$) to the
strong coupling regime ($g_{1D}\rightarrow \pm \infty $) via Feshbach
resonances and confinement-induced  resonances \cite{Olshanii}. 
$g_{1D}>0$ ($<0$) is the contact repulsive (attractive) interaction.
Usually,  one  defines the interaction strength as $c=mg_{1D}/\hbar ^{2}$ and dimensionless parameter 
$\gamma =c/n$ for convenience. 
Here we denoted the total density $n=n_\uparrow+n_\downarrow$, the magnetization $M=(n_\uparrow-n_\downarrow)/2$, 
and the polarization $P=(n_\uparrow-n_\downarrow)/n$, where $n=N/L$ is the linear density and $L$  is the length of the system.

The model was solved by Bethe ansatz  \cite{Yang:1967,Gaudin:1967}, where the Bethe ansatz  wave numbers $\left\{k_i\right\}$  
are the quasimomenta of  fermions satisfying the Bethe ansatz equations 
\begin{eqnarray}
& &\exp(\mathrm{i}k_jL)=\prod^M_{\ell = 1}
\frac{k_j-\Lambda_\ell+\mathrm{i}\, c/2}{k_j-\Lambda_\ell-\mathrm{i}\, c/2},\nonumber\\
& &\prod^N_{\ell = 1}\frac{\Lambda_{\alpha}-k_{\ell}+\mathrm{i}\,
c/2}{\Lambda_{\alpha}-k_{\ell}-\mathrm{i}\, c/2}
 = - {\prod^M_{ \beta = 1} }
\frac{\Lambda_{\alpha}-\Lambda_{\beta} +\mathrm{i}\,
c}{\Lambda_{\alpha}-\Lambda_{\beta} -\mathrm{i}\, c} . \label{BE}
\end{eqnarray}
Here $j = 1,\ldots, N$ and $\alpha = 1,\ldots, M$, with $M$ spin-down fermions.
The parameters $\left\{\Lambda_{\alpha}\right\}$ are the
rapidities for the internal spin degrees of freedom.
The energies of the ground state and all excited states are given
in terms of these quasimomenta, i.e. $E=\frac{\hbar
^2}{2m}\sum_{j=1}^Nk_j^2$.

\begin{figure}[t]
{{\includegraphics [width=1.0\linewidth,angle=-0]{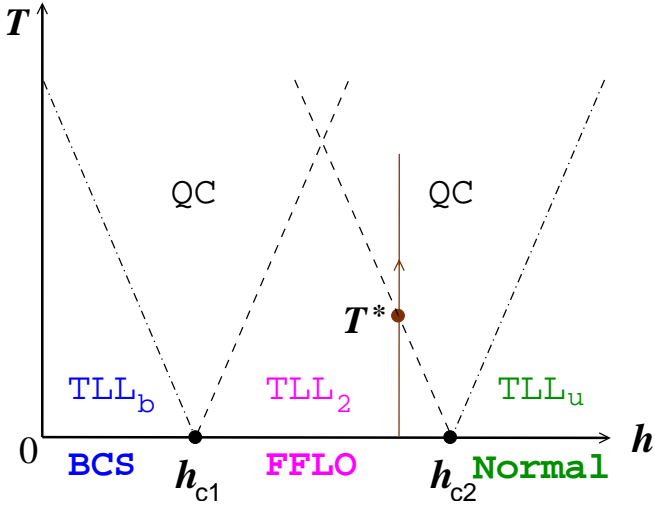}}}
\caption{Schematic phase diagram in $T-H$ plane \cite{Zhao}. 
Three quantum phases at zero temperature: BCS pairs with zero polarization, FFLO with polarization and normal Fermi gas. 
  At finite temperatures the crossover temperatures $T^*$ separate the free fermion quantum criticality from the LLs, i.e. the LL of bound pairs $LL_b$, two-component LL of pairs and excess fermions $LL_2$ and the LL of free fermions $LL_u$, respectively. 
  The critical exponents associated with critical points are uniquely determined by the symmetry of the system, discussion in detail see \cite{Guan-Ho,Zhao}.  Figure extracted from \cite{Zhao}.}
  \label{fig:QC-Fermi}
\end{figure}

In the attractive interacting regime, at zero temperature, the  quasimomenta $k_i$ of two  atoms with different spin states form two-body bound states,
i.e., $k_j=\Lambda_j\pm   \mathrm{i} \frac{1}{2} c$, whereas the momenta of excess fermions are  real in quasimomentum space. 
The distributions of these quaismomenta comprise the complexity of many-body physics.
At the zero temperature and under an  external magnetic field, the attractive Fermi gas have three different ground states: 
a fully-paired phase for the external field is less than the lower critical field $H_{c1}$, a fully-polarized ferromagnetic phase  for the external field is greater than the upper crtical field $H_{c2}$, and  the  significant  Fulde-Ferrel-Larkin-Ovchinnikov (FFLO) \cite{Fulde,Larkin}  like  pairing phase   for a intermedium field $H_{c1}<H<H_{c2}$. 
In the grand canonical ensemble,  the phase diagram in the $T-H$ plane  presents  a  universality class of criticality of the 1D interacting fermions, see   Figure \ref{fig:QC-Fermi}.
Experimental measurement of phase diagram of two-component ultracold ${}^6$Li atoms trapped in an array of 1D tubes through the finite temperature density profiles confirms theoretical predictions \cite{Liao}.
The experimental developments on studying this one-dimensional interacting fermions \cite{Moritz:2005,Zurn:2012,Wenz:2013} advance our understanding of many-body physics from integrability.

\begin{figure}[t]
{{\includegraphics [width=0.90\linewidth,angle=-0]{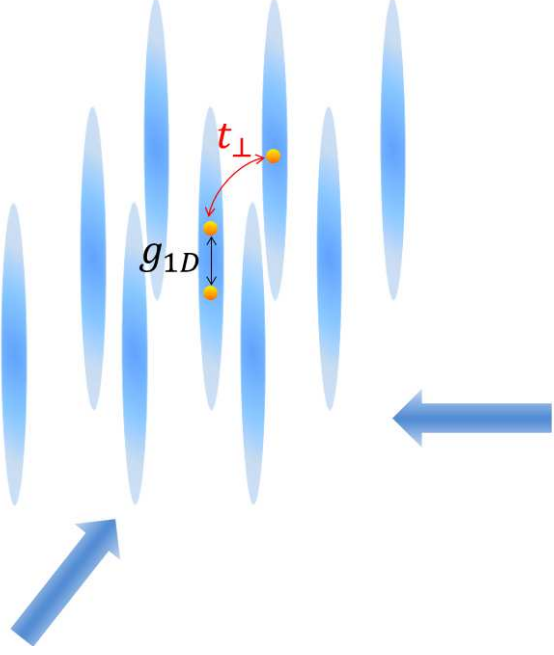}}}
\caption{Schematic quasi-one-dimensional system with particle tunneling between tubes. Each  tube can  be treated as a 1D man-body system in grand canonical ensemble. The FFLO-like  phase can exist at finite temperatures  for  a none zero tunneling strength, also see  \cite{Bogoliubov}.  }
  \label{fig:QC-3D}
\end{figure}

At zero temperature, the FFLO correlations  \cite{Feiguin:2007,Tezuka:2008,Rizzi:2008a,Luscher:2008,Batrouni:2008,Baur:2010,Parish:2007,Liu:2008b,Zhao:2008,Edge:2009,Datta:2009,Pei:2010,Devreese:2011,Kajala:2011,Chen:2012,Bolech2012,Lu2012}  are  the major concern from theory and experiment. 
From Bethe ansatz point of view, the low energy excitations correspond to changes in particle numbers (pairs or unpaired fermions) close to Fermi points. 
As a consequence, the system exhibits local scale invariance, i.e. conformal invariance.
Thus various correlation functions can be analytically calculated using conformal field theory \cite{Lee:2011a,Schlottmann2012,LGSB}, for example 
the one particle Green's function,
\begin{eqnarray}
&& G_{\uparrow}(x,t) =
\langle\psi_{\uparrow}^{\dagger}(x,t)\psi_{\uparrow}(0,0)\rangle
\\ &&\approx \frac{A_{\uparrow,1}\cos\left(\pi(n_{\uparrow}-2n_{\downarrow})x\right)}
{|x+\mathrm{i}v_{u}t|^{\theta_{1}}|x+\mathrm{i}v_{b}t|^{\theta_{2}}}
+\frac{A_{\uparrow,2}\cos\left(\pi
n_{\downarrow}x\right)}{|x+\mathrm{i}v_{u}t|^{\theta_{3}}|x+\mathrm{i}v_{b}t|^{\theta_{4}}},\nonumber
\end{eqnarray}
the charge density correlation function
$G_{nn}(x,t)$ 
\begin{eqnarray}
&&G_{nn}(x,t) = \langle n(x,t)n(0,0)\rangle \\
&&\approx \nonumber
n^{2}+\frac{A_{nn,1}\cos\left(2\pi(n_{\uparrow}-n_{\downarrow})x\right)}{|x+\mathrm{i}v_{u}t|^{\theta_{1}}}
+\frac{A_{nn,2}\cos\left(2\pi
n_{\downarrow}x\right)}{|x+\mathrm{i}v_{b}t|^{\theta_{2}}},
\end{eqnarray}
and the pair correlation   $G_{p}(x,t)$
\begin{eqnarray}
&&G_{p}(x,t) =
\langle\psi_{\uparrow}^{\dagger}(x,t)\psi_{\downarrow}^{\dagger}(x,t)\psi_{\uparrow}(0,0)\psi_{\downarrow}(0,0)\rangle
\\ &&\approx
\frac{A_{p,1}\cos\left(\pi(n_{\uparrow}-n_{\downarrow})x\right)}{|x+\mathrm{i}v_{u}t|^{\theta_{1}}|x+\mathrm{i}v_{b}t|^{\theta_{2}}}
+\frac{A_{p,2}\cos\left(\pi(n_{\uparrow}-3n_{\downarrow})x\right)}{|x+\mathrm{i}v_{u}t|^{\theta_{3}}|x+\mathrm{i}v_{b}t|^{\theta_{4}}}.\nonumber
\end{eqnarray}
In the above correlation functions, for strong coupling regime, the exponents  can be given explicitly in \cite{Lee:2011a,Schlottmann2012}. 
We observe  that the leading order for the long distance asymptotics of the pair
correlation function $G_{p}(x,t)$  oscillates with wave number
$\Delta k_F$, where $\Delta k_F =\pi(n_{\uparrow}-n_{\downarrow})$.
The oscillation in pair correlation  is  caused by an
imbalance in the densities of spin-up and spin-down fermions, i.e.,
$n_{\uparrow}-n_{\downarrow}$, which gives rise to a mismatch in
Fermi surfaces between both species of fermions. 
This  spatial
oscillation shares a similar signature as the Larkin-Ovchinikov (LO)
pairing phase \cite{Larkin}.
This can be regarded as the existence of a quasi-long range order. 
At finite temperatures, these correlations decay  exponentially in 1D systems.
The average distance between the pairs is the same order as between the unpaired fermions.
The pairs lose dominating nature and the critical temperature is $T=0$. 
Therefore the 1D analog of FFLO spacial oscillation nature in the pair correlation function is unable to probe through the trapped Fermi gas at finite temperatures. 
Nevertheless a quasi-1D systems that the fermions can tunneling from one tube to another would give rise to a finite critical temperature  for the existence of the ordered phase at finite temperatures \cite{Bogoliubov}, see Fig.~\ref{fig:QC-3D}.

At finite temperatures,   full  thermodynamics of the model is determined by the  TBA equations \cite{Takahashi71-74} that give rise to the universal quantum criticality. 
In the grand canonical ensemble,  the grand partition function
 $Z={\mathrm {tr}} (\mathrm{e}^{-\cal{H}/T})=\mathrm{Exp}({-G/T})$ is written  in terms of the Gibbs
free energy $G = E - HM^z - \mu n - TS$  with respect to the magnetic field $H$, 
chemical potential $\mu$ and entropy $S$.   
At finite temperatures, the density distribution functions of pairs, unpaired fermions and spin strings involve the densities of 
`particles' $\rho_r(k)$ and `holes' $\rho_r^h(k)$, here  $r=1,2$ for single excess  fermions and bound pairs.  
The dressed energies,  $\epsilon^{\rm b}(k) := T\ln( \rho_2^h(k)/\rho_2(k) )$ and 
$\epsilon^{\rm u}(k) := T\ln( \rho_1^h(k)/\rho_1 (k) )$  characterize the excitation energites  for paired and unpaired fermions.
The equilibrium states are determined by the minimization condition of
the Gibbs free energy that gives rise to the TBA equations
\begin{eqnarray}
\epsilon^{\rm
  b}(k)&=&2(k^2-\mu-\frac14{c^2})+Ta_2*\ln(1+\mathrm{e}^{-\epsilon^{\rm b}(k)/T} )
  \nonumber\\
& &+ \, Ta_1*\ln(1+\mathrm{e}^{-\epsilon^{\rm u}(k)/{T}})\nonumber\\
\epsilon^{\rm
  u}(k)&=&k^2-\mu-\frac12{H}+Ta_1*\ln(1+\mathrm{e}^{-\epsilon^{\rm b}(k)/{T}})\nonumber\\
& &-T\sum_{n=1}^{\infty}a_n*\ln(1+\eta_n^{-1}(k))\nonumber\\
\ln
  \eta_n(\lambda)&=&\frac{nH}{T}+a_n*\ln(1+\mathrm{e}^{-\epsilon^{\rm u}(\lambda)/{T}})\nonumber
\\&&+\sum_{n=1}^{\infty}T_{nm}*\ln(1+\eta^{-1}_m(\lambda)).\label{TBA-Full}
\end{eqnarray}
The function $\eta_n(\lambda) := \xi^h_n(\lambda)/\xi_n (\lambda ) $ is the ratio of the
string densities.  
The same mathematical notations as used in previous spin-1 Bose gas. 
The function $T_{nm}(\lambda)$ is given in \cite{Takahashi-b}. 
The Gibbs free energy per unit length is given by $G=-p^b-p^u$ where the effective pressures of the unpaired fermions and  bound pairs are given by
\begin{equation}
p^r=\frac{rT}{2\pi}\int_{-\infty}^{\infty}dk\ln(1+\mathrm{e}^{-\epsilon^{\rm r}(k)/{T}}) \nonumber
\end{equation}
with $r=1$ for unpaired fermions and $r=2$ for paired fermions.   
The TBA equations (\ref{TBA-Full}) indicate that the band energies of the bound pair and single particle depend on the changes of the pressures of  bound pairs and excess fermions. 
The function $a_n(x)$ originated from the two-body scattering amplitude  encodes interaction effect. 
In these equations with  proper expansion, one can obtain the similar  first- and second-order Landau coefficients  which reveal the forward scattering phase shift like the quasiparticle energies in  Fermi liquid \cite{WangYP}.
 Later we shall briefly introduce  such connection to the Wilson ratio in  the attractive Fermi gas. 

\begin{figure}[t]
{{\includegraphics [width=1.0\linewidth,angle=-0]{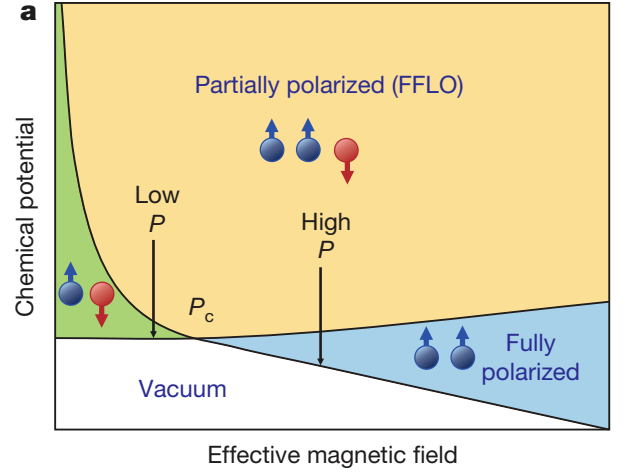}}}
\caption{Phase diagram in $\mu-H$ plane.  The solid lines show the numerical result from the TBA equations (\ref{TBA-Full}) in $T=0$ limit  \cite{Liao}, also see \cite{Guan-Ho}. The system has a partially polarized core surrounding by either fully paired or fully polarized wings in an harmonic trap, which is in agreement with this theoretical prediction.
 Orso \cite{Orso} for the first time presented this phase diagram using the Bethe ansatz equations, similar study was presented in \cite{HuiHu,Guan2007}. Figure extracted from \cite{Liao}. }
\label{fig:Phase_diagram}
\end{figure}

The $T=0$ phase diagram  has been worked out by Orso \cite{Orso}  and by others \cite{HuiHu,Guan2007}  using BA equations, which describe the ground state within a canonical ensemble.
In terms of the dimensionless quantities   
$\tilde{\mu}:= \mu/\varepsilon_{b}$, $h:= H/\varepsilon_{b}$, $t:= T/\varepsilon_{b}$ and 
$\tilde{n}:= n/|c|=\gamma^{-1}$, where $\varepsilon_b = \hbar^{2} c^2/(4m)$ is the binding energy,  
the phase diagram is shown in Fig.~\ref{fig:Phase_diagram}.  
It consists of three  phases: fully paired phase $P$, ferromagnetic phase $ F$, and
partially paired $ PP$ or ($FFLO$-like) phase. 
The system has spin population imbalance caused by a difference in the number of spin-up and spin-down atoms. 
These critical phase boundaries have been discussed in  \cite{Guan-Ho}.

\begin{figure}[t]
\includegraphics[width=1.0\linewidth]{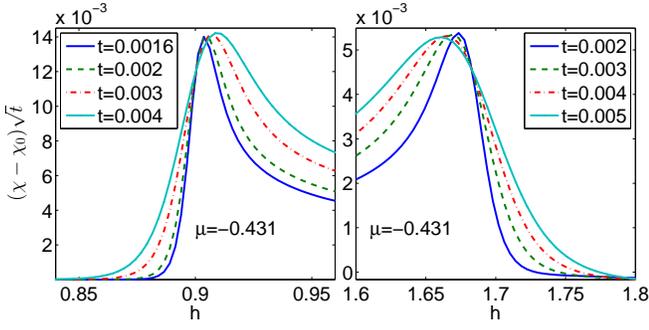}
\caption{ (Color online) The temperature scaled susceptibility  vs magnetic field $h$ at different temperatures for a fixed values of $\tilde{\mu}=-0.431$. The susceptibility  curves at different temperatures intersect at the lower  critical point  $H_{c1}$ (left panel) and the upper critical point  $H_{c2}$ (right panel) that signifies  the scaling form (\ref{sus}) with dynamical critical exponent $z=2$ and correlation length exponent $\nu =1/2$. 
} 
\label{fig:sus-fermion}
\end{figure}

At low  temperatures and strong coupling regime, the equation of states for the strongly attractive gas  has been obtained    by Guan and Ho in  \cite{Guan-Ho}.
 The analytical expression of the total pressure $p=p^b+p^u$  was presented for the regime $\left| \gamma \right| \gg 1$ and $t
\ll 1$ \cite{Guan-Ho}. 
The dimensionless pressure $\tilde{p}=
\tilde{p}^{b}+\tilde{p}^{u}$ of the system was  found to be  \cite{Guan-Ho}
\begin{eqnarray}
\tilde{p}^{b} &=&-\frac{t^{3/2}f_{3/2}^{b}}{2\sqrt{\pi }}\left( 1-\frac{
t^{3/2}f_{3/2}^{b}}{16\sqrt{\pi }}-\frac{t^{3/2}f_{3/2}^{u}}{\sqrt{2\pi }}
\right)   \nonumber \\
\tilde{p}^{u} &=&-\frac{t^{3/2}f_{3/2}^{u}}{2\sqrt{2\pi }}\left( 1-\frac{
t^{3/2}f_{3/2}^{b}}{\sqrt{\pi }}\right)   \label{pressure-fermion}
\end{eqnarray}%
where 
\begin{eqnarray}
A_{b}&=&2\tilde{\mu}+1-\tilde{p}^{b}-4\tilde{p}^{u}-\frac{t^{5/2}f_{5/2}^{b}}{
16\sqrt{\pi }}-\sqrt{\frac{2}{\pi }}t^{5/2}f_{5/2}^{u}\nonumber  \\
A_{u}&=&\tilde{\mu}+\frac{h}{2}-2\tilde{p}^{b}-\frac{t^{5/2}f_{5/2}^{b}}{2\sqrt{\pi
}}+f_{s} \label{chemical2}\\
f_{n}^{b}&=&\mathrm{Li}_{n}\left(-e^{A_{b}/t}\right), \,\,\,f_{n}^{u}=
\mathrm{Li}_{n}\left(-e^{A_{u}/t}\right). 
\end{eqnarray}
Here 
\begin{eqnarray}
f_{s}&=&te^{-h/t}e^{-2\tilde{p}^{u}/t}I_{0}\left( 2\tilde{p}
^{u}/t\right),\nonumber\\
I_{n}\left( x\right) &=&\sum_{k=0}^{\infty }\frac{1}{
k!\left( n+k\right) !}\left( x/2\right) ^{n+2k}.\nonumber
\end{eqnarray}
This term  indicates that  the $SU(2)$ spin degree of freedom ferromagnetically couples to the unpaired Fermi sea. 
In fact,   this  spin wave contribution to the function $A_u$   is negligible due to an exponentially small contributions at low temperatures.
 Although this equation of state (\ref{pressure-fermion}) was derived in the regime for strong attraction and low temperatures, it presents  a  new theoretical scheme to treat quantum criticality of 1D  ultracold  Fermi  gases with arbitrary interaction strength  by employing analytical methods.  
 The key observation is that the  dimensionless form of the pressure of the system 
 \begin{equation}
 \tilde{p}(t, \tilde{\mu}, h)\equiv p/(|c|\varepsilon_b)=\tilde{p}^b+\tilde{p}^{u},
 \end{equation}
can be expressed as a universal scaling form near quantum phase transitions 
\begin{eqnarray} 
\tilde{p}(t, \tilde{\mu}, h)=\tilde{p}_0 +T^{\frac{d}{z}+1}{\cal P}\left(\frac{\mu-\mu_c}{T^{\frac{1}{\nu z}}},\frac{H-H_c}{T^{\frac{1}{\nu z}}}\right).\label{scaling-p}
\end{eqnarray}
 Where $\tilde{p}_0$  is the background pressure before quantum phase transition. In the second term $ {\cal{P}}$ is a dimensionless scaling function. 
 The critical fields $\mu_c$ and $H_c$ present the phase boundaries in Fig.~\ref{fig:Phase_diagram} \cite{Guan-Ho}.
 In the above equations, $\tilde{p}^{b,u}$ are  the dimensionless pressures of bound pairs and single fermions. 
 The existence of the scaling form of Eq. (\ref{scaling-p})    illustrates the microscopic origin of 
quantum criticality and provide analytic insight into continuum field theory that  describes universal scaling theory  in the vicinities of critical points.  
Thus the thermodynamical properties can be cast into universal scaling forms  such as density and compressibility as presented in   Eqs. (\ref{uni_scal_dens}) and  (\ref{uni_scal_kappa}). 
The density, compressibility, magnetization and susceptibility have the following  scaling forms 
\begin{eqnarray}
n(\mu,H, T)&=&n_0+T^{\alpha}{\cal G}\left(\frac{\mu-\mu_c}{T^{\frac{1}{\nu z}}},\frac{H-H_c}{T^{\frac{1}{\nu z}}}\right),
\label{uni_scal_dens-2}\\
\kappa(\mu,H, T)&=&\kappa_0+T^{\beta }{\cal F}\left(\frac{\mu-\mu_c}{T^{\frac{1}{\nu z}}},\frac{H-H_c}{T^{\frac{1}{\nu z}}} \right), \label{uni_scal_kappa-2}\\
M(\mu,H,T)&=&M_0+T^{\alpha }{\cal K}\left(\frac{\mu-\mu_c}{T^{\frac{1}{\nu z}}},\frac{H-H_c}{T^{\frac{1}{\nu z}}}\right),
\label{uni_scal_m}\\
\chi(\mu,H,T)&=&\chi_0+T^{\beta }{\cal O}\left(\frac{\mu-\mu_c}{T^{\frac{1}{\nu z}}},\frac{H-H_c}{T^{\frac{1}{\nu z}}}\right)
\label{uni_scal_chi}
\end{eqnarray}
with the exponents $\alpha =\frac{d}{z}+1-\frac{1}{\nu z}$ and $\beta =\frac{d}{z}+1-\frac{2}{\nu z}$. The scaling functions
${\cal{G}}(x),\,{\cal{F}}(x),\, {\cal{K}}(x)$ and ${\cal{O}}(x)$ can be worked  out explicitly  from the TBA equations (\ref{TBA-Full})  in the region $T> |H-H_c|, |\mu-\mu_c|$.
From the equation of state (\ref{pressure-fermion}), one can analytically  examine quantum criticality of the model in an harmonic trap \cite{Yin2011}.   
Here we demonstrate that 
the dimensionless susceptibility $\tilde{\chi}=\chi \varepsilon_b /|c|$   near the two critical points, i.e., in  the vicinity of the critical points $h_{c1}$ and $h_{c2}$,  presents a universal scaling form 
\begin{equation}
\chi \sim \frac{|c|}{\epsilon_{\rm b}}\left[\lambda_0+\lambda_s t^{\frac{d}{z}+1-\frac{2}{\nu z}} {\rm Li}_{-\frac{1}{2}}  
\left( -\mathrm{e}^{\frac{\alpha_s(h-h_{c})}{t^{\frac{1}{\nu z}}}  } \right)\right]. \label{sus}
\end{equation}
For strong attraction and  near the lower critical point $h_{c1}=-2\tilde{\mu} +\frac{32}{3\pi\sqrt{2}}(\tilde{\mu} +1/2)^{3/2}$, there is  no background susceptibility, i.e. $\lambda_0=0$. 
The constant  $\lambda_s\approx \frac{1}{8\sqrt{2}\pi }\left(1-\frac{6}{\pi}\sqrt{(h-h_{c1})/2} \right)$ with $\alpha_s=1/2$.
In the above equations we used the dimensionless units  $t=T/\epsilon_b$ and $h=H/\epsilon_b$. 
The scaling form (\ref{sus}) reads off the universality class of quantum criticality of the dynamical critical exponent $z=2$ and correlation length exponent $\nu =1/2$.   
We plot this scaling law of susceptibility near two critical points for fixed values of magnetic field in  Fig.~\ref{fig:sus-fermion}. 
In contrast,  near the upper critical point  $h_{c2} \approx 1+(3\pi)^{2/3}(\tilde{\mu} +1/2)^{3/2}-2(\tilde{\mu}+1/2)$, there is a  background susceptibility.
The universal form of susceptibility given in (\ref{sus}) with the following  constants 
\begin{eqnarray}
\lambda_0&\approx &1/( 8\sqrt{2}\pi \sqrt{\lambda_2^{u}}),\qquad \lambda_s\approx  \lambda_2^u/(\pi^2\sqrt{\pi} ),\nonumber \\
\lambda_2^u&\approx &\left( 3\sqrt{2} \pi(2\tilde{\mu} +1)/8 \right)^{2/3} -16 \left(\tilde{\mu}+1/2\right)^{3/2}/(3\sqrt{2}\pi ),\nonumber\\
\alpha &\approx& \frac{1}{\sqrt{2} \pi}\left(3\sqrt{2}\pi (2\tilde{\mu} +1) \right)^{1/3}\nonumber
\end{eqnarray}
maps out the quantum criticality with universal dynamical critical exponent $z=2$ and correlation length exponent $\nu =1/2$ for the phase transition from the FFLO to the normal free Fermi gas.

\subsection{Wilson ratio in one dimension}

In the scenario of universal low energy physics, Landau Fermi liquid theory \cite{Landau:1956,Landau:1957,Landau:1958} describes universal  matter of quasiparticles in typical  electronic metals, Kondo problem  and He-3 liquid etc. 
The fermion excited outside the Fermi surface adiabatically evolves into  a quaisparticle with the same charge, spin and momentum. 
These quasiparticles at the Fermi energy  have  an infinite lifetime. 
Even for these systems with  strong interaction,  such quasiparticle excitations still resemble the free fermions with a renomalized excitation energy { $\omega(k)=\frac{\hbar k_F}{m*}(k-k_F)+\cdots$}. 
Here $m^{*}$ is the effective mass of quasiparticles and $k_F$ is the Femi momentum. 
Consequently,  the effective magnetic moment and the mass of the quasiparticle are renomalized in terms of Landau parameters  during the scattering process close to the surface. 
 For  forward scattering process,  interaction just changes effective mass in the density of state.  
The specific heat increases linearly with the temperature $T$ as
\begin{eqnarray}
 c_v=\frac{1}{3}\frac{m^*k_Fk_B^2T}{\hbar^3 }
\end{eqnarray}
because  only the electrons  
within $k_BT$ near the Fermi surface contribute to the specific heat. 
The susceptibility is independent of temperature since only the electrons within $\mu_B g H$  
near the Fermi surface contribute to the magnetization. More  explicitly,
\begin{equation}
\chi =\frac{m^*k_F}{\pi^2\hbar}\frac{\mu_F^2g^2}{1+F^a_0}. 
\end{equation}
In the above equations $F^a_0$ is the Landau parameter. 
We see that the  specific heat and susceptibility   only depend on the density of state and Landau parameters in forward scattering process.
This universal nature is captured by a dimensionless constant, i.e. 
the Wilson ratio \cite{Sommerfeld,Wilson:1975}, which is defined as the ratio of the magnetic susceptibility $\chi$  to specific heat $c_v$ divided by temperature $T$,
\begin{equation}
R_W=\frac{4}{3}\left(\frac{\pi k_B}{\mu_B g }\right)^2\frac{\chi}{c_v/T}
\label{ratio}
\end{equation}
The effective mass was  canceled out  in this  ratio. 
Thus  the Wilson  ratio is a constant at the renormalization fixed point of the interacting fermionic systems, despite it involves two thermodynamic properties. 
For example, most weakly correlated metals this ratio is a unity  $ R_W=1$.
It was proved that by Wilson \cite{Wilson:1975} $R_W=2$ in the Kondo magnetic impurity problem in low temperatures. 

In contrast, in 1D many-body systems  the quasiparticle description breaks down due to the fact that all particles participate in the low energy physics.
 The main reason for this  is that the elementary excitations in
 1D are still eigenstates. 
Therefore one  can not find a simple operator,
acting on the ground state, to get an individual  quasiparticle excitation, unlike that in 
higher dimensional  Fermi liquid.
The low-lying excitations  in 1D actually  form  collective motion of bosons, i.e., the LL.  
In  this regard, one can say ``{\em there will be no Fermi liquid in one dimension}" \cite{Georges}.
However, the 1D system is critical in the gapless phase of LL, which is regarded as the critical phase at the renormalization fixed point.
In such a low energy sector, elementary excitations close to the Fermi point in 1D systems  essentially determine the critical behaviour and universal thermodynamics of the LL.
Despite a breakdown of the quasiparticle description in 1D,  the Fermi liquid nature can be retained in  low energy sector \cite{Schulz,WangYP,Johnston}. 
Typical phenomenon of spin-charge separation in 1D interacting system  is described by the effective Hamiltonian
\begin{eqnarray}
H=H_c+H_{\sigma}+\frac{2g_1}{(2\pi \alpha)^2}\int dx \cos \left(\sqrt{8\phi_{\sigma}}\right),
\end{eqnarray} 
where the effective Hamiltonians in spin and charge are given by  
\begin{equation}
H_{\nu}=\int dx \left(\frac{\pi v_{\nu} K_{\nu}}{2} \Pi^2_\nu+\frac{v_{\nu}}{2\pi K_{\nu} }\left(\partial _x \phi_{\nu} \right)^2\right)
\end{equation} 
for $\nu=c, \, \sigma$. 
Here $\phi_{\nu}$ and $\Pi_{\nu}$ obey the standard Bose commutation relations $\left[ \phi_\nu, \Pi_\mu \right]=\mathrm{i} \delta _{\nu\mu}\delta(x-y)$.  
The interaction term $g_1$ characterizes the backscattering process.  The $K_{c,\sigma}$ are the Luttinger parameters  and $v_{c,\sigma}$  are the sound velocities in  charge and spin sectors, respectively. 
 In term of spin-charge separation, the Wilson ratio at the fixed renormalization point reads
\begin{equation}
R_W=\frac{2v_c}{v_c+v_\sigma}.
\end{equation}
This naturally suggests  that for noninteracting systems the ratio is unity and there does not exist  spin-charge separation mechanism. 
However, for a strong repulsion, the spin transportation  tends to zero so that the Wilson ratio tends  to $2$. 
This signature was numerically confirmed for  the 1D Hubbard model by the Bethe ansatz solution \cite{Johnston}.
It  was recently pointed out \cite{GYFBLL} that  the Wilson ratio   quantifying   interaction effects  and spin fluctuations presents the  universal nature of quantum liquids for  both Fermi liquid and Luttinger liquid. 
Beyond the spin-charge separation scenario, the  two important features of the Fermi liquid are retained for  the renornalaization fixed point phase in 1D Fermi gases  on a general ground, namely the specific heat is linearly proportional to temperature whereas the susceptibility is independent of temperature.
Nevertheless,  exact Bethe ansatz solutions would provide an alternative but a precise way to capture  the nature  of Fermi liquid.
This study provides an  intrinsic connection between the Fermi liquid and the LL.

In the context of cold atoms, the effective magnetic field $H$ depends  on the chemical potential bias $H=\mu_{\uparrow}-\mu_{\downarrow}$. 
The magnetization depends on the difference between the spin-up and -down fermion densities  $\Delta n=2M^z=n_{\uparrow}-n_{\downarrow}$.
In general,  for the attractive interaction, the  homogeneous system is described by two coupled Fermi
gases of bound pairs and excess fermions in the charge sector and
ferromagnetic spin-spin interaction ordering  in the spin sector.
At low temperatures, the ferromagnetic spin waves contributions to the low energy physics are marginal. 
In contrast to the spin-charge separation in repulsive Fermi gas, only the charge density fluctuations dominate the low energy physics for the attractive Fermi gas. 
In order to work out  the susceptibility  universality,  we  define two effective chemical potentials for bound pairs and excess fermions
\begin{equation}
\mu_b=\mu+\varepsilon_b/2,\qquad \mu_u= \mu +H/2.
\end{equation}
Where the total chemical potential is given by $\mu =(\mu_{\uparrow} +\mu_{\downarrow})/2$ and $\varepsilon_b$  is the binding energy.
For fixed total density, by definition,  the  susceptibility can be written in term of two charge susceptibilities  
\begin{eqnarray}
\chi&=&\frac{1}{2}\frac{\partial \Delta n}{\partial \Delta \mu}=1/ \left( \frac{1}{\chi_u}+\frac{1}{\chi_b}\right).\label{susceptibility}
\end{eqnarray}
Where  the charge susceptibilities of bound pairs and excess fermions were defined  by  $\chi _{\rm b,u}=\frac{1}{2} \partial n_{\rm b,u}/\partial \mu_{\rm b,u}|_{\mu_{u,b}} $. In the above equations, $n_b$ and $n_u$ are  the densities of pairs and excess fermions. 
This relation is general and independent of integrability. 
 This means that  the two physical processes, i.e.  the break  of pairs and alignment of the spins,  occur paralelly.  
 The effective susceptibilities for the renormalization fixed point  LL of bound pairs and the LL of excess fermions are expressed as 
\begin{eqnarray} 
\chi_{\rm b}=1/(\hbar \pi v_N^{b})\quad \chi_{\rm u}=1/(4\hbar \pi v_N^{{\rm u}}).
\end{eqnarray}
 The  density stiffness parameters are obtained from 
$v_N^r=\frac{L}{\pi\, \hbar }\frac{\partial ^2E_0^{r}}{\partial N_r^2}$ for a Galilean invariant system, 
with $r=1$ for excess fermions and $r=2$ for bound pairs. 
For the strongly interacting regime ($\gamma >1$), these density stiffness are given by \cite{GYFBLL}
\begin{eqnarray}
\left\{\begin{array}{l}
v_{N}^{{\rm b}}= \frac{\hbar \pi n_2}{4m}\left(1+\frac{4}{|c|}(n-3n_2)+\frac{3}{c^2}(4n^2-24nn_2+30n_2^2)\right)\\
v_N^{{\rm u}}=\frac{\hbar\pi n_1 }{m}\left( 1+\frac{4}{|c|}(n-2n_1)+\frac{4}{c^2}(2n^2+10n_1^2-12nn_1)   \right) \end{array}
\right.  \nonumber
\end{eqnarray}
Here $n_1$ and $n_2$ are the density of excess fermions and pairs, respectively.


For $T=0$ the TBA dressed energies have  Fermi points  at $Q_{\alpha}^{\pm}=\pm Q_{\alpha}$ with $\alpha=1,2$ for excess fermions and bound pairs, thus 
the Fermi velocities of unpaired fermions
and bound pairs are defined  as
\begin{equation}
v_{\alpha} =\pm\frac{\varepsilon'_{\alpha}(\pm
Q_{\alpha})}{p'_{\alpha}(Q_{\alpha})}=\pm\frac{\varepsilon'_{\alpha}(\pm
Q_{\alpha})}{2\pi\rho_{\alpha}(\pm Q_{\alpha})}
\end{equation}
with $\alpha =1,2$.
Where $\rho_{\alpha}$ are the density distribution functions for excess fermions and pairs.
Thus the finite-size
corrections to the ground state energy and finite temperature corrections to the free energy \cite{Guan2007,Zhao}
\begin{eqnarray}
E&=&E_{0}^{\infty}-\frac{\pi C }{6L^{2}} \left( v_{1} +v_2 \right)\nonumber \\
F&=& E_0^{\infty}-\frac{\pi C T^2}{6}\left(\frac{1}{v_{1}}+\frac{1}{v_2}\right)\label{finite-E}
\end{eqnarray}
indicates a universal critical scaling of conformal field theory with central change $C=1$  \cite{Affleck:1986}. 
In the above equations $E_0^{\infty}$ is the ground state energy in thermodynamic limit.  
The  exact
expressions for the velocities can be found
from the relation $v_s^{r} =\sqrt{\frac{L}{mn}\frac{\partial ^2E_0^{r}}{\partial L^2}}$. 
These expressions 
 (\ref{finite-E}) are universal in regard to the low energy sector and valid for arbitrary interaction regime.
The low-lying excitations
are decoupled into two massless  excitations within Fermi seas of bound pairs and single excess fermions which are
described by two Gaussian theories. 

In the strong coupling limit,  the velocities are given by 
\begin{equation}
v_s^{r}=\frac{\hbar }{2m}\frac{2\pi n_r }{r}  \left(1+ \frac{2A_r}{|c|} +\frac{3A_{r}^2}{c^2} \right)
\end{equation}
with $r=1,2$. Here  $A_1=4n_2$ and $A_2=2n_1+n_2$. 
Consequently, the specific heat  for the two-component LL is given by 
\begin{eqnarray}
c_v&=&\frac{\pi C T}{3}\left(\frac{1}{v_{1}}+\frac{1}{v_2}\right). \label{cv}
\end{eqnarray}
This result is valid for arbitrary interaction strength. 
This  linear $T$-dependence of the specific heat is a characteristic of the two-component LL  with linear dispersions in branches of pairs and single fermions. 
%
%
%
Deviation from this universal LL specific heat (\ref{cv})  indicates a breakdown of the  linear temperature-dependent relation.
This naturally defines the  crossover temperature $T^*$ which charaterizes  
a universal crossover from a relativistic  dispersion into a  nonrelativistic  dispersion \cite{Maeda2007,Zhao}.  

Moreover, from the LL free energy (\ref{finite-E}) one can check that the susceptibility is temperature independent in the phase of the two-component LL. 
The separation signature of the susceptibility (\ref{susceptibility}) and specific heat (\ref{cv})  naturally suggests that the two branches of gapless excitations in the 1D FFLO-like phase form two collective LLs. Thus 
the low energy  (long wavelength) physics of the strongly attractive Fermi gas  is  described by an effective Hamiltonian
\begin{eqnarray}
H_{\rm eff} &=&\sum_{\sigma ={\rm u,b }} \int dx \left[ \frac{\pi v_{\sigma }}{2} \Pi_{\sigma}^2+\frac{v_{\sigma} }{2\pi K^{(\sigma)}  } (\partial _x \phi_{\sigma} )^2\right]\nonumber\\
&&-\frac{H}{2}\int dx \frac{\partial_x \phi_{\rm u} }{\sqrt{\pi} }-\mu \int dx \frac{\partial_x \phi_{\rm u} +2\partial _x \phi_{\rm b} }{\sqrt{\pi} }\label{Eff-H}
\end{eqnarray}
as long as the spin fluctuation is frozen out, see a discussion in \cite{Vekua}. 
Here the fields $\partial_x \phi_{\sigma},\Pi=\frac{1}{\pi }\partial_x \theta_{\sigma} $ with $\sigma={\rm b,u}$ are the density and current fluctuations for the pairs and unpaired fermions.  In this case, the current-current interaction becomes irrelevant in the gapless phase.  
Thus the low energy physics of the FFLO-like phase is described by a renormalization fixed point of the two-component TTL class, where the interaction effect enters into the above collective velocities, or equally speaking that  the effective masses of the two LLs are varied by the interaction.

\begin{figure}[htbp]
\includegraphics[width=1.0\linewidth]{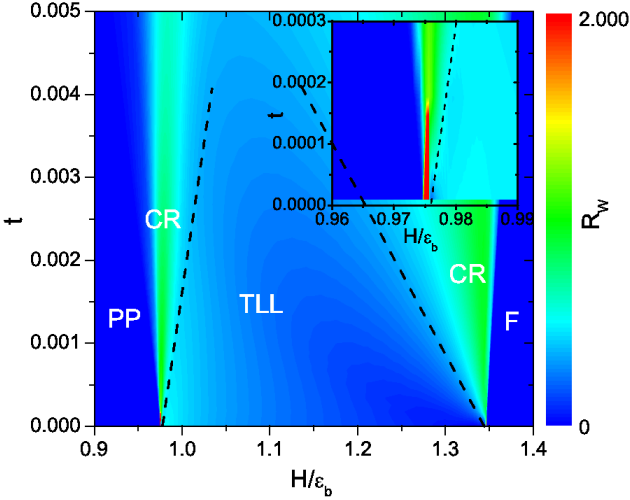}
\caption{(Color online) Low temperature  Wilson ratio $R_W$ (\ref{ratio}) of the attractive Fermi gas with  dimensionless interaction $|\gamma| =10$ \cite{GYFBLL}.
Eq. (\ref{W_R}) provides a criterion for  the two-component LL phase in the region below the dashed lines. The derivation from this formula 
 indicates the crossover  temperature $T^*\sim |H-H_c|$ (dashed line) separating the relativistic liquid  from the nonrelativistic liquid. 
$R_W =0$ for both the LL  of pairs (PP) and the LL of excess fermions (F). In the critical regimes (CR) $R_W$ gives a 
temperature-dependent scaling. However, at  the two critical points, the ratio reveals anomalous enhancement, see the inset.  Figure extracted from  \cite{GYFBLL}. } 
\label{fig:contour}
\end{figure}


The linear temperature-dependent nature of the specific heat and the temperature-independent  susceptibility retain  the important features of  the Fermi liquid.
The Wilson ratio of  the attractive Fermi gas is given by 
\begin{equation}
R_W=\frac{4}{\left( v_N^{\rm b}+4v_N^{\rm  u}\right)\left(\frac{1}{v_s^{\rm b}}+\frac{1}{v_s^{\rm  u}}\right)}\label{W_R}
\end{equation} 
for  the two-component LL phase in the attractive Fermi gas. 
This simple relation valid for a wide range of 1D spin-1/2 interacting fermionic systems.
This result is universal in terms of the density stiffness $v_N^{{\rm b,u}}$ and sound velocity $v_s^{{\rm b,u}}$ for 
pairs b and excess single fermions u. 
Fig.~\ref{fig:contour} shows that at finite temperatures  the contour plot of $R_W$ can map out not only  the two-component LL 
phase but also  the quantum criticality of the attractive Fermi gas.  
It is worth noting that  Wilson ratio for the 1D attractive Fermi gas  is significantly different from the ratio obtained 
for the field-induced gapless phase in the quasi-1D gapped spin ladder \cite{Ninios2012}, 
where the gapless phase is a single-component LL \cite{Schulz1991,WangYP}.
Again, deviation from the Wilson ratio (\ref{W_R}) gives the crossover temperature $T^*\sim |H-H_{c}|$ separating   the LL from   the free fermion liquid near the critical points.   
In contrast to the phenomenological LL parameter, the Wilson ratio provides a testable parameter for quantifying universal nature of quantum liquids of  interacting fermions in one, two and three dimensions.
%

\begin{figure}[htp]
\includegraphics[width=1.0\linewidth]{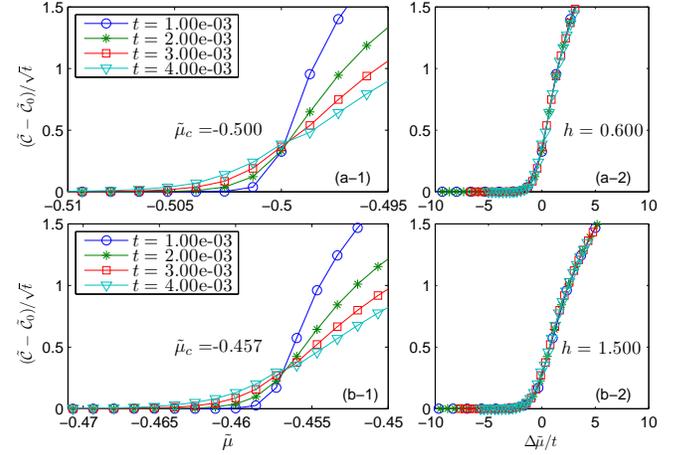}
\caption{(Color online)  Fig.~ (a-1) and (b-1) show the temperature rescaled dimensionless Contact of the attractive Fermi gas vs chemical potential  for different temperatures   near phase boundaries  $V-P$  and $F-PP$, respectively. Correspondingly, the  figure  (a-2) and (b-2) show the  rescaled Tan's contact vs $\tilde{\mu}$ ($|\tilde{\mu} -\tilde{\mu}_c |/t$) at different temperatures collapse into one line near  the critical points. The point of intersection identifies the phase boundaries and the critical exponents  from the Contact at finite temperatures. Figure extracted from  \cite{CJZG}.  }
\label{fig:Contact}
\end{figure}

\subsection{Critical behaviour of Contact  in one dimension}

Universal relations which hold independently of the interaction strength always attract  great deal of attention from theory and experiment.
The universal quantity  called contact which  has  been  for the first time found in unitary Fermi gases by Tan \cite{Tan} provides new insight into two-body correlations in short distance limit, i.e. $\langle \hat{n}({\bf x})\hat{n}({\bf x+d})\rangle\sim {C}|{\bf d}|/(4\pi)$ with $d\to 0$, where  $\hat{n}({\bf x})$ is the density operator at position ${\bf x}$ and  ${C}$ is the contact and. 
Tan's contact, which  measures the two-body correlation at short distances in dilute systems, also remarkably presents  the universality of many-body systems of ultra cold atoms  \cite{CJZG}.  
It  connects various properties of the system, ranging from  the  tail of  large momentum distribution  $n_\sigma (k) \rightarrow \mathcal{C}/k^4$ as $k\to \infty$, thermodynamics $P=\frac{2}{3}\varepsilon +\hbar^2\mathcal{C}/(12\pi m a)$,  high frequency dynamic structure factor and adiabatic relation with the changes in the scattering length $\left(\frac{dE}{da^{-1}} \right)_s=-\frac{\hbar^2}{4\pi m}\mathcal{C}$, see \cite{Braaten,Hu:2011,Doggen:2013,Smith:2014,Yan:2013,Partridge:2005}.
Here  $a$ is the scattering length. 
The significant feature of  Contact is that it can be applied to any many-body system of interacting bosons and fermions in 1D, 2D and 3D and exists in normal phase and superfluid phase.
Tan's contact  leads to immediate implications in cold atomic  physics \cite{Stewart:2010,Sagi:2012,Hoinka:2013}. 

The study of Tan's relation in 1D many-body systems provides essential information on local pair correlations \cite{Barth:2011}. 
Recent study  on  the universality  of Contact for the spin-1/2 Fermi gas shows that Contact  remarkably connects  quantum criticality  of  many-body systems \cite{CJZG}. 
Tan's adiabatic  contact in 1D is defined $\mathcal{C} =\frac{4g_{1D}^2m^2}{\hbar^4} \int dR \langle \psi^{\dagger }_{\uparrow}\psi^{\dagger }_{\downarrow}\psi_{\downarrow}\psi_{\uparrow}(R)\rangle$ \cite{Barth:2011}. 
In practice,  it reads 
\begin{equation}
\mathcal{C}= \frac{2m}{\hbar^2}\left(\frac{d E}{d (-a_{1D})}\right)_{s}
\end{equation}
It  connects various thermodynamical properties of 1D many-body systems  via the general thermodynamical relations
\begin{eqnarray}
d p =n d\mu +s dT +m_z dH-\mathcal{C} d (a_{1D}). 
\end{eqnarray}
In the above equations $E$ is the energy eigenvalue, $s$ is the entropy and $m_z$ is the magnetization. 
This relation provides insightful connections of Contact  to thermodynamical  properties $n,\,s,\,m_z$ via 
\begin{eqnarray}
\left(\frac{\partial \mathcal{C} }{\partial \mu }\right)_{T,H,a_{1D}}&=&\left(\frac{\partial n}{ \partial (-a_{1D})}  \right)_{\mu, T, H},\nonumber\\
\left(\frac{\partial \mathcal{C} }{\partial H }\right)_{\mu, T,a_{1D}}&=&\left(\frac{\partial m_z}{ \partial (-a_{1D})}  \right)_{\mu, T, H},\nonumber\\
\left(\frac{\partial \mathcal{C} }{\partial T }\right)_{\mu,H,a_{1D}}&=&\left(\frac{\partial s}{ \partial (-a_{1D})}  \right)_{\mu, T, H}.\label{C-d}
\end{eqnarray}
With the help of Eq. (\ref{pressure-fermion}) we  can prove  the  novel existence of universal scaling behaviour of the Contact.
 In grand canonical ensemble,  the contact is given    by 
\begin{eqnarray}
 \mathcal{C} \equiv \frac{c^2}{2} G^{(2)} =-\frac{c^2}{2} \left(\frac{\partial p}{\partial c}\right)_{\mu,H,T}.\label{contact-p}
 \end{eqnarray} 
Here $G^{(2)}$ is the total local pair correlation and $p$ is the pressure per unit length.  

From the scaling form of the pressure (\ref{scaling-p}), we see that critical fields $H_c$ and $\mu_c$ essentially depend on the interaction.
So that the contact defined by (\ref{contact-p}) must possess scaling behaviour like other thermodynamical properties such as  density, compressibility, magnetization and susceptibility. 
It was found \cite{CJZG} that the  universal scaling forms of  the contact and its derivative $\partial \mathcal{C} /\partial\mu$ with respect to the driven parameter $\tilde{\mu}$ (or ${h}$) read 
\begin{eqnarray}
\tilde{\mathcal{C} }^{(2)}&=&\mathcal{C}_{0}+\lambda_GT^{(d/z)+1-(1/vz)}\mathcal{F}(\frac{\mu-\mu_{c}}{T^{1/vz}}) \label{C-form}\\
\frac{\partial \tilde{ \mathcal{C}} }{\partial\tilde{\mu}}&=&\tau_{0}+\lambda_DT^{(d/z)+1-(2/vz)}\mathcal{G} \left(\frac{\mu-\mu_{c}}{T^{1/vz}}\right), \label{C-dd}
\end{eqnarray}
where $\mathcal{C}_{0}$ ($\tau_0$) is the background  contact (derivative of the contact) which is  temperature independent. 
The constant  $\lambda_{G,D}$  depend on the critical values of $\mu_c$ and $H_c$.  
In the above equations    $\mathcal{F}(x)$ and  $\mathcal{G}(x)$  are  the dimensionless  scaling functions.
The critical phenomena of   Contact (\ref{C-dd}) are  valid for full interaction strength. 
The finite temperature  Contact  shows that  the scaling form (\ref{C-form}) with the dimension $d=1$ for  the model read off  the dynamic exponent $z=2$ and the correlation exponent $v=1/2$ \cite{CJZG}.
Fig.~\ref{fig:Contact} shows the  universal scaling behaviour  for  the phase transitions from vacuum $V$  to the fully paired phase $P$  and from fully-polarized phase $F$ to the partially-polarized phase $PP$    through  the numerical solution of the TBA equations (\ref{TBA-Full}). 
The scaling behaviour   of the  derivative of Contact with respect to the driving parameter was  demonstrated  in connections to various physical properties, see \cite{CJZG}.
This finding sheds light on the universal feature of Tan's contact  in connection to   quantum criticality  in low and higher dimensions. 
In particular, it provides insights into the study of  critical behaviour of the Contact near the critical temperature  for  the unitary Fermi gas \cite{Sagi:2012}.


\section{V. Outlook}

\begin{table}[t]
\begin{ruledtabular}
 \caption{Three integrable points for spin-3/2 quantum gases \label{table1}}
 \begin{tabular}{c|c ccccc  l}
 No. & $g_{00}$ & $g_{2,2}$ & $g_{2,1}$ & $g_{2,0}$ & $g_{2,-1}$  & $g_{2,-2}$ \\
 \hline
 (i) & $c$ & $c$ & $c$ & $c$ & $c$ & $c$ \\
 (ii) & $3c$ & $c$ & $c$ & $c$ & $c$ & $c$ & \\
 (iii) & $c$ & $-c$ & $c$ & $-c$ & $c$ & $-c$ & \\
 \end{tabular}
\end{ruledtabular}
\end{table}

Quantum criticality  provide  a promising way to explore a broad range of novel phenomena and  hidden symmetries  of  quantum many-body systems, for example,   the 1D  quantum Ising model with both a transverse field and a longitudinal field displays the largest exceptional group  E8 symmetry \cite{E8,Saleur}.
A  pedagogical review of  theoretical results for the Ising model universality class, including $E_8$  
S-matrix, correlation functions, integrability breaking, etc, was discussed in terms of integrable field theory and critical phenomena \cite{Delfino:2004}.
This hidden E8 symmetry was dramatically   demonstrated in experiment  \cite{Coldea:2010,Morris:2014}.  
In this scenario  experiments on quantum criticality of ultracold atoms \cite{Gemelke:2009,Huang:2010,Huang:2011,Zhang:2012} 
  open up a new way to explore  critical phenomena  in quantum  many-body systems.  
Here it is appropriate   to quote  the  insightful  perspective on  quantum criticality remarked by Coleman and Schoefield \cite{Coleman}: ``Quantum criticality may be a highly effective catalyst for the formation of new stable types of material behaviour, providing an important new route for the design and discovery of new classes of material Ó.

Moreover, integrability with cold atoms offers  a precise method  to treat quantum critical phenomena in many-body systems of bosons and fermions, see review \cite{Guan-RMP}.
Along this line,  it  is highly desirable to study quantum criticality of large spin systems with high symmetries  in  this exact manner.  
The large spin systems of cold atoms exhibit  rich internal structures that  may  result in new quantum phases and multi-component LLs \cite{Ho-Yi,Yi-Ho,Wu:2003,Honerkamp:2004,Gorshkov:2010,Cazalilla3}.
In this regard, the spin-3/2 fermionic systems are likely to provide  an ideal model towards a precise understanding of large spin non-$SU(N)$ symmetries.

The simplest high spin Hamiltonian \cite{Wu:2005,Lecheminant:2005,Controzzi,Jiang:2009}
\begin{eqnarray}
 \label{H} \hat H = -\sum_{j=1}^N \frac{\partial^2}{\partial x_j^2}
 +\sum_{i\neq j}^N \sum_{lm} g_{lm} \hat P^{lm}_{ij}\delta(x_i-x_j)  \label{Ham-h}
\end{eqnarray}
describes dilute spin-3/2 atomic gases of $N$ fermions with contact
interaction constrained by periodic boundary conditions to a line of length
$L$. 
Here the summation $\sum_{lm}$ is carried out for $l=0,2$ and
$m=-l,-l+1, \cdots, l$.
In the above equations  projection operator $\hat{P}^{lm}_{ij}=|lm\rangle\langle lm|$ projects the total
spin-$l$ state onto  spin-$m$ in the $z$-direction of two colliding atoms $i$
and $j$.
The interaction strength in the channel
$|lm\rangle\langle lm|$ is given by  $g_{lm}=-2\hbar^2/(ma_{1D}^{lm})$. 
It is remarkable that the model exhibits different spin symmetries $SU(4)$, $SO(5)$ and $SO(4)$ while the integrability is preserved \cite{Jiang:2014}, see the Table I. 
This model has rich spin pairing states, i.e. spin-$J$ pairs with $J=0,\,2$ which lead to    the Fulde-Ferrel-Larkin-Ovchinnikov
like pair correlations and multi-component LLs for all  interaction strength. 
These spin-$J$ pairs form  two subspaces,  with interaction strength $c>0$ and $c<0$ respectively 
\begin{eqnarray}
\left\{ \begin{array}{l}  \hat\phi_{2,0}=
 (\hat\psi^\dag_{-3/2}\hat\psi^\dag_{3/2},
 -\hat\psi^\dag_{1/2}\hat\psi^\dag_{-1/2})/\sqrt2,\\
 \hat\phi_{2,2}=\hat\psi^\dag_{1/2}\hat\psi^\dag_{3/2},\,\,
 \hat\phi_{2,-2}=\hat\psi^\dag_{-3/2}\hat\psi^\dag_{-1/2},\end{array} \right. \nonumber \\
 \,\,\,\,
 \left\{ \begin{array}{l} \hat\phi_{0,0}
 =(\hat\psi^\dag_{-3/2}\hat\psi^\dag_{3/2}
 +\hat\psi^\dag_{1/2}\hat\psi^\dag_{-1/2})/{\sqrt2},\\
 \hat\phi_{2,1}=\hat\psi^\dag_{-1/2}\hat\psi^\dag_{3/2},\,\,
 \hat\phi_{2,-1}=\hat\psi^\dag_{-3/2}\hat\psi^\dag_{1/2}.\end{array}\right.\label{pair-J}
\end{eqnarray}

\begin{figure}[t]
\includegraphics[width=\linewidth]{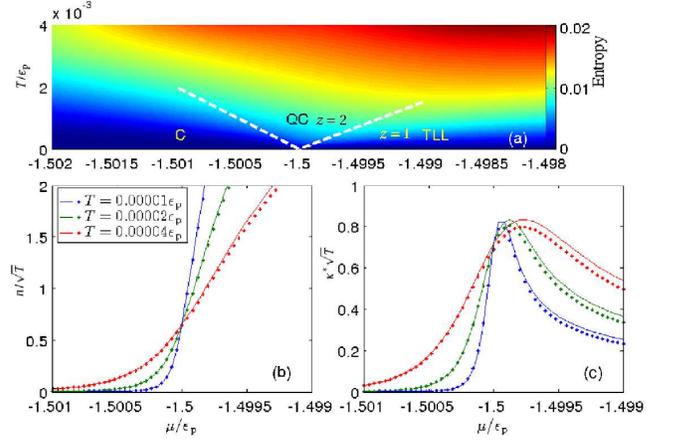}
\caption{\label{fig:high}   (a) Quantum criticality: entropy vs chemical potential $\mu$  near  the phase transition  V-FFLO   for  $h=\epsilon_{\rm p}$ \cite{Jiang:2014}.
(b)  and (c) show the universal scaling forms of (\ref{QC-high}) with $\mu_c=-1.5\epsilon_b$. 
 We denote  $C$ as  the classical region, $QC$ as  the quantum critical regime while  $LL$ stands for  theTomonaga-Luttinger liquid.  The white dashed lines indicate the crossover temperature.   The  good agreement between the analytical result (solid lines) and   the numerical solution (dotted  lines) further confirms the quantum criticality (\ref{QC-high}). Figure extracted from \cite{Jiang:2014}.}
\end{figure}

It is very interesting that the model exhibits FFLO-like  pair
correlation function in a mixed phase of $\hat\phi_{2,2}$ pairs and single fermions of $|{3/2}\rangle$ atoms 
\begin{eqnarray}
\langle G|\hat\phi_{2,2}(x,t)\hat\phi_{2,2}(0,0)|G\rangle\approx
 \frac{A_0\cos(\pi \Delta k_F x)}
 {|x+{\rm i} v_{\rm u}t|^{\theta_{\rm u}}
  |x+{\rm i} v_{\rm p}t|^{\theta_{\rm p}}},
\end{eqnarray}
where $\theta_{\rm u}=1/2$, $\theta_{\rm p}=1/2+n_{\rm p}/c$ and $n_{\rm u,p}$
are the densities of unpaired $|{3/2}\rangle$ atoms and atomic pairs
$\hat\phi_{2,2}$, respectively.
The quantum criticality of the model for the phase transition from the vacuum into the FFLO phase is governed by the 
 scaling functions 
\begin{eqnarray}
\left\{  \begin{array}{l} \frac{n}{\sqrt{T}}\approx \frac{1}{2\sqrt{\pi}}
 \left[F_{-\frac12}\Big(\frac{\mu-\mu_c}{T}\Big)
  +2^{\frac32}    F_{-\frac12}\Big(\frac{2(\mu-\mu_c)}{T}\Big)\right]\\
 \kappa^*\sqrt{T}\approx \frac{1}{2\sqrt{\pi}}
 \left[F_{-\frac32}\Big(\frac{\mu-\mu_c}{T}\Big)
  +2^{\frac52}    F_{-\frac32}\Big(\frac{2(\mu-\mu_c)}{T}\Big)\right] \end{array} \right. \label{QC-high}
\end{eqnarray}
where $F_\beta(x)$ is  the  Fermi-Dirac function and $\mu_c=-1.5\epsilon_{\rm p}$. 
Indeed the scaling functions of density  and compressibility (\ref{QC-high})  identify the dynamic critical exponent $z=2$ and correlation length exponent $\nu=1/2$, see the Fig.~\ref{fig:high}.
This preliminary  study of quantum criticality of large spin fermionic systems opens a way to treat high spin phenomena in an analytical fashion. 
It is particularly interesting to probe  Wilson ratio and Tan's contact with quantum criticality of    large spin systems.

Moreover,  recent  experimental developments  with  ultracold  atoms provide exciting opportunities to test quantum dynamics of many-body systems. 
In particular, the quantum dynamics of mobile spin  impurity  give deep  understanding of the motion of impurity and polaronic behaviour \cite{Fukuhara1}. 
Microscopic observation of two magnon bound states helps one to understand quantum walks of magnons predicted by the  Bethe ansatz solvable model \cite{Fukuhara2,Natan:2014}. 
This research suggests  further investigation of  quantum quench dynamics of 1D interacting bosons and fermions.
  Essentially, in the 1D Lieb-Liniger gas of bosons,  two types of excitations--Bogoliubov phonons and type II excitations,  give rise to significantly different dynamics.
The type II excitation is regarded as the quantum dark soliton \cite{Sato:2012}.
 Creating  such type II excitations in the  trapped gas  would demonstrate unique 1D quantum dynamics  through  a 2D array of  1D tubes. 
  Quantum quench dynamics of interacting fermions and bosons in 1D  play an essential role in understanding  nonequilibrium  evolution of  isolated systems  beyond the usual thermal Gibbs mechanism \cite{Rigol:2007}.
 The  generalized Gibbs ensemble \cite{Rigol:2008,Caux:2013,Nardis,Natan2014,Natan2013}  is  used  in discussions of   non-thermal distributions in  isolated systems with conserved laws.
 Recently there have been several papers focusing on the study of quench dynamics in regard to  the validity of the  generalized Gibbs Ensemble (GGE), see   
 \cite{arXiv1,arXiv2,arXiv3,arXiv4,arXiv5}, etc.  
 This research has been becoming a new frontier in cold atoms and condensed matter physics. 
 However, understanding thermalization of isolated many-body systems still imposes a  big theoretical  challenge.

{\bf Acknowledgment.} 
The author  thanks   Tony Guttmann, Murray T  Batchelor,  Helen Perk, Angela Foerster, Jean-Sebastien Caux, Tetsuo Deguchi  and zoran ristivojevic  for discussions and help with proofreading. 
This  work has been supported by  the National Basic Research Program of China under Grant No. 2012CB922101 and NNSFC under grant numbers 11374331. He   has been partially supported by the Australian Research Council. 
 He acknowledges the Beijing Computational Science Research Center and KITPC,  Beijing  for their kind hospitality.

\end{document}